\documentclass[twocolumn,floatfix,showkeys,titlepage,balancelastpage,raggedbottom,amsfonts,amssymb,amsmath]{revtex4}
\usepackage{epsfig}
\usepackage{bm}
\usepackage{times}
\usepackage{mathptmx}
\usepackage{amsmath}
\usepackage{graphicx}
\usepackage{graphics}
\usepackage{amssymb}
\usepackage{multirow}
\usepackage[english]{babel}
\begin{document}
\title{Random packing of small blocks: 
pressure effects, orientational correlations and application to polymer-based composites}
\author{Danilo Sergi}
\affiliation{University of Applied Sciences (SUPSI), 
The iCIMSI Research Institute, 
Galleria 2, CH-6928 Manno, Switzerland}
\author{Claudio D'Angelo}
\affiliation{University of Applied Sciences (SUPSI), 
The iCIMSI Research Institute, 
Galleria 2, CH-6928 Manno, Switzerland}
\author{Giulio Scocchi}
\affiliation{University of Applied Sciences (SUPSI), 
The iCIMSI Research Institute, 
Galleria 2, CH-6928 Manno, Switzerland}
\author{Alberto Ortona}
\affiliation{University of Applied Sciences (SUPSI), 
The iCIMSI Research Institute, 
Galleria 2, CH-6928 Manno, Switzerland}
\date{\today}
\begin{abstract}
Packing is a complex phenomenon of prominence in many natural and industrial processes 
(liquid crystals, granular materials, infiltration, melting, flow, sintering, segregation, 
sedimentation, compaction, etc.). A variety of computational methods is available in particular 
for spheroid particles. Our aim is to apply the principle of the random sequential addition algorithm but
with small blocks of varying size and orientation. Here the main purpose is to reproduce the observed arrangement 
of graphitic assemblies in polymeric matrices. Random packing is improved by applying an external pressure implemented 
with a drifted diffusive motion of the fillers. Attention is also paid to the emergence of structural 
and orientational order. Interestingly, mixtures of fillers of irregular shapes can be dealt 
with efficiently using the proposed algorithm. 
\end{abstract}
\maketitle

\section{Introduction}

\begin{figure}[b]
\includegraphics[width=7.225cm]{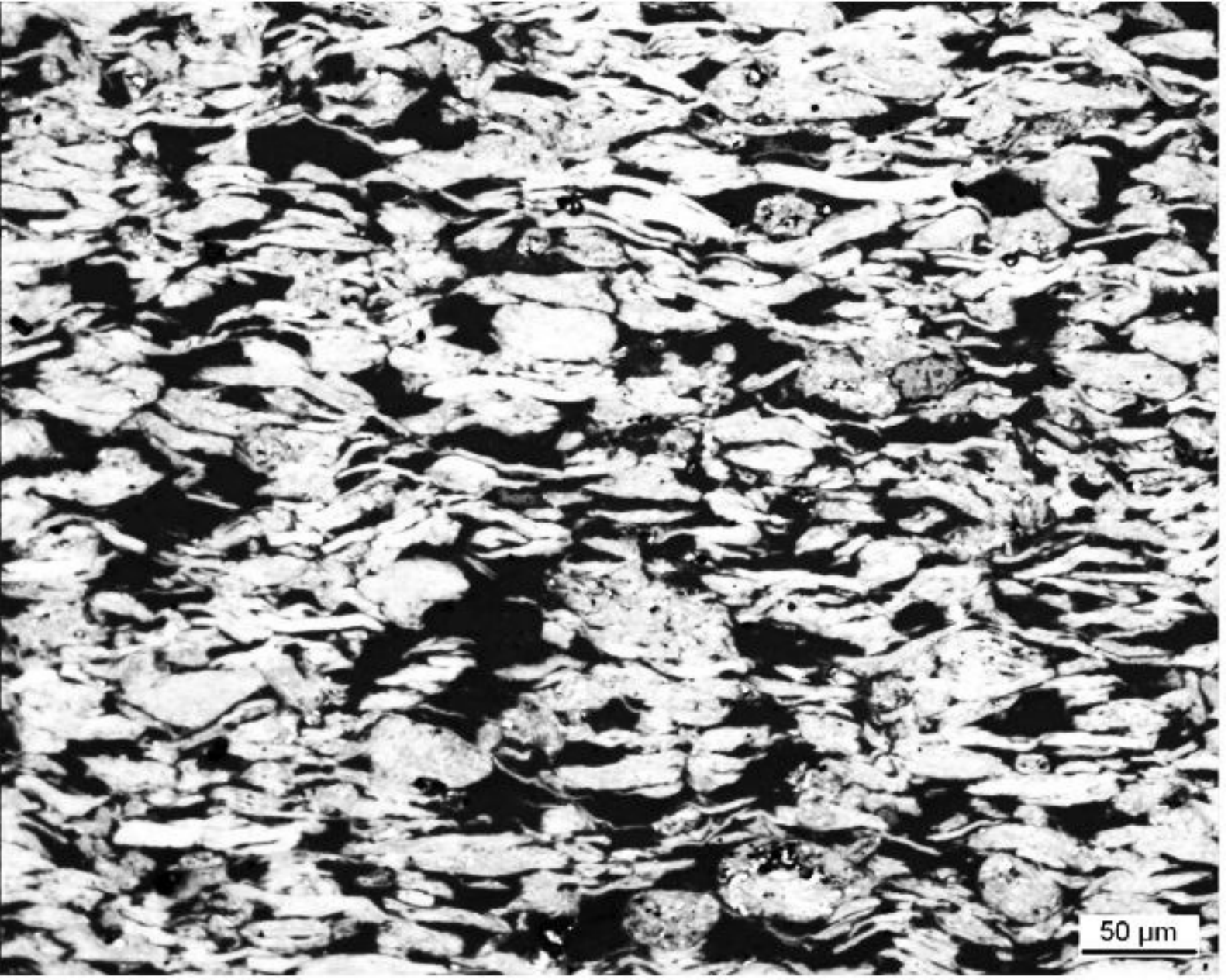}
\caption{Microstructure geometry of a typical polymeric matrix reinforced with a graphitic powder. 
It appears that the graphitic phase (lighter) displays to a certain extent rectangular 
characteristics. By $2$D image processing it is determined a volume fraction of around $62\%$.
\label{fig:micro}}
\end{figure}

In the present Article we are faced with the problem of realizing the densest
packing with small blocks out of a bimodal size distribution. The motivation
is to generate the arrangement of graphitic assemblies in polymeric matrices as
obtained after compaction (see Fig.~\ref{fig:micro}). In these materials, the graphitic
phase can reach high volume fractions. It follows that the
packing structures of these systems demand attention, especially when the 
electrical properties are under investigation, since the polymer is insulating
\cite{tunnel,tunnel2}. More generally, the polymeric matrix is enriched with a
second dispersed phase in order to enhance a wide range of properties of the
resulting composite material (mechanical, electrical and thermal) \cite{composites}. 
In the appearance, this kind of materials display quite disordered microstructures 
that nonetheless reveal attributes of key importance in several aspects \cite{book_torquato}. 
At the extremes, random packing can be addressed with detailed, fully deterministic 
numerical methods like classical molecular dynamics \cite{gb,re2} or with fully random processes 
\cite{widom,sherwood}. In between, dynamic approaches allowing the particles to change their
size \cite{stillinger,science_princeton,jodrey,part_part} appear to be the most established. 
Typically, it arises that the shape of the fillers has a significant influence on the 
packing properties (see in particular Refs.~\cite{science_princeton,prl_princeton}). 
Unfortunately, the mathematical treatment of general morphologies is far from being 
trivial and for this reason most of the research has so far concentrated on 
ellipsoidal particles (spheres, oblates and prolates), and in a few exceptions on
cubes \cite{cube1,cube2,dpd_cube} and others \cite{digipack}. 
Our work is based on the random sequential addition (RSA) algorithm \cite{widom} 
with the possibility to use hard fillers of arbitrary shape, similarly to what 
reported in Refs.~\cite{williams1,williams2,williams3} for $2$D, but with coarser discretization for 
efficiency concerns. In order to favor compaction, we introduce the effect of an external pressure 
via a biased diffusive motion of the fillers. In principle, blocks can pack very well because 
they have extended planar surfaces \cite{science_princeton}. When they can be rotated, severe geometrical 
constraints are imposed in their neighborhood. For this reason, the structural properties of the final configurations 
are analyzed in detail by means of radial distribution functions.
Furthermore, we work out a method to detect orientational correlations at a local
level. Our results suggest that orientational correlations have a range that can go beyond that of
structural order.


\section{Random addition strategy}

The domain to be filled with small, impenetrable blocks has all side lengths equal to $1$
with periodic boundary conditions. We build the blocks from small spheres
of radius $r$, for the reasons explained in Ref.~\cite{williams1}. The blocks are obtained 
as follows. A single sphere
is first placed randomly in the domain with uniform distribution. This
sphere is that with the smallest $x$, $y$ and $z$ coordinates composing
the block (before rotation). The other spheres are added so as to be aligned 
along the three cartesian axes with their centers separated by the distance $r$.
Let $l_{x}$, $l_{y}$ and $l_{z}$ be the side lengths of the block; for example,
the length of the $x$ side is $l_{x}=(n_{x}+1)r$, where $n_{x}$ is the number of 
spheres aligned along this direction. The size of the block is fixed randomly by choosing
$n_{x}$, $n_{y}$ and $n_{z}$ from
three Gauss distributions of means $\langle n_{x}\rangle=(\langle l_{x}\rangle-r)/r$, 
$\langle n_{y}\rangle$ and $\langle n_{z}\rangle$; the three distributions have the same variance
$\sigma^{2}$ and subsequent computations are done using the integer parts of $n_{x}$, $n_{y}$ and $n_{z}$. 
The resulting block is then rotated around the $z$, $x$ and $y$ axes, in this order, by the 
angles $\theta_{z}$, $\theta_{x}$ and $\theta_{y}$, respectively. These three angles of 
rotation are chosen randomly within a given interval from a uniform distribution. For the 
rotation, the axes are translated with their origin in the center of the first sphere.

Let $\Delta x$, $\Delta y$ and $\Delta z$ be the distances in absolute value
along the three axes between the centers of two spheres. This means that the two
centers are separated by $\sqrt{(\Delta x)^{2}+(\Delta y)^{2}+(\Delta z)^{2}}$.
The block is added to the domain if for every center of its spheres at least 
one $\Delta$ is larger than $2r+d_{\mathrm{min}}$ \cite{note}. 
If this condition is not fulfilled there is 
overlap: the block is discarded and the above procedure is repeated.

We want to verify relatively fast the overlap condition. To this end,
the domain is divided into five layers of equal width parallel to the
$xy$ plane. The centers of the spheres are stored in five lists, one
for every layer. In order to avoid overlap at the interfaces, the spheres
falling within a distance of $2r$ from an interface are also added to the
list of the adjacent layer. In this way, the overlap condition is checked
only among spheres belonging to the same list.  
\begin{figure}[t]
\includegraphics[width=7.225cm]{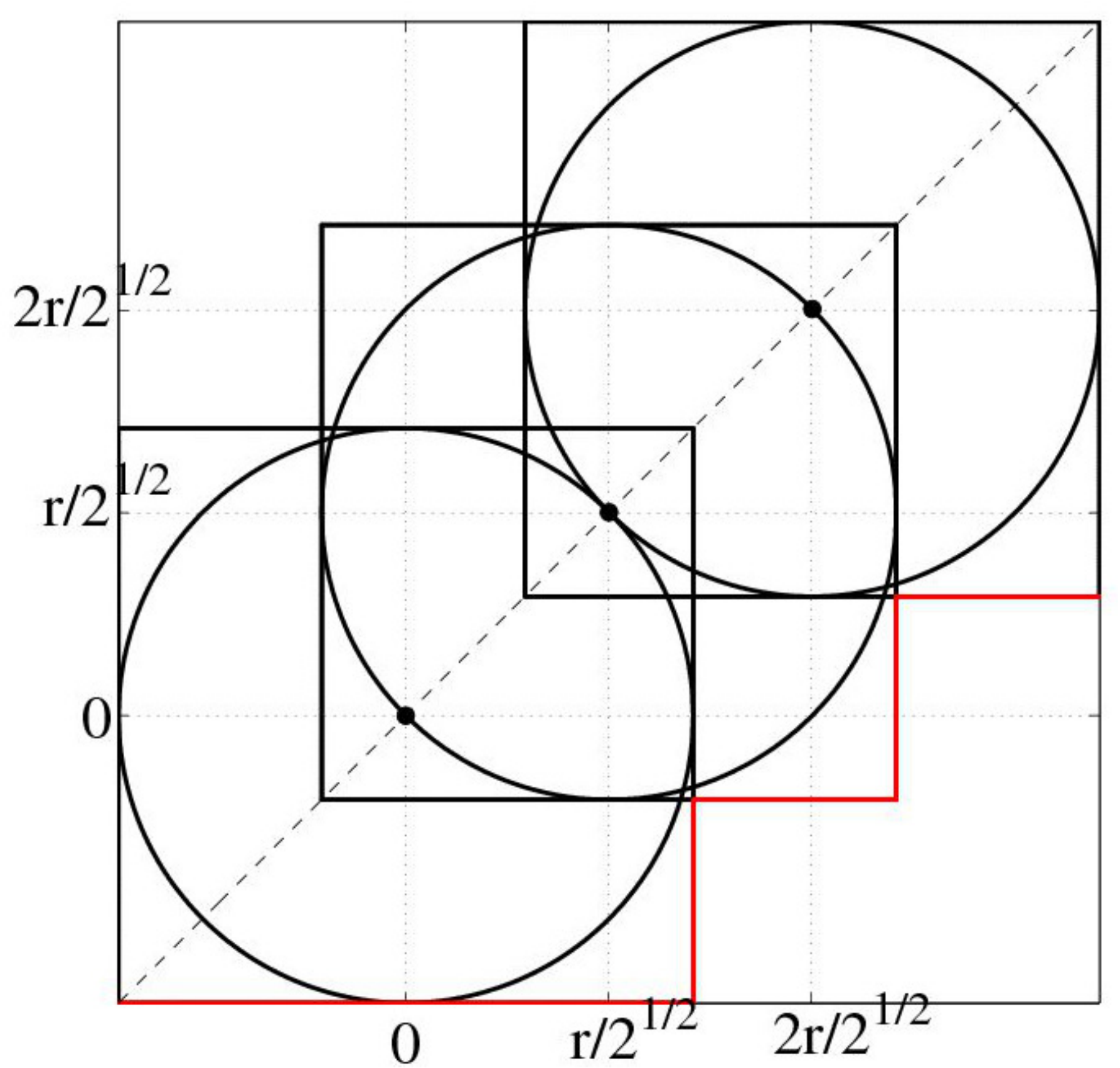}
\caption{How the spheres are inscribed in the cubes. The origin of axes is in the center of the
first placed sphere of the block. In red is shown the boundary of the block in the $xy$ plane
in the absence of tilt.
\label{fig:rough}}
\end{figure}

The binning of the domain is regular, leading to $N^{3}$ voxels. 
Every side of the domain is thus divided into $N$ bins.
Knowing the centers of the spheres, the real blocks are readily obtained at the end
of the simulation by inscribing every sphere in a cube of side length $2r$.  
Every voxel inside the cubes will be considered as 
occupied by the filler phase. Voxels outside the cube will be considered instead as occupied 
by the matrix phase. By this procedure, the surface of the resulting
parallelepipeds is rough. The degree of roughness is arguably in line
with experimental observations. Indeed, we find that the
depth of the defects is given approximately by the formula 
$r\mid \cos\theta_{z}\sin\theta_{z}\mid$. The roughness is
of course maximal for $\theta_{z}=\pi/4$.
By simple geometrical arguments, we find that the depth
of the incidents at the surface is approximately at most $r/2$ (see Fig.~\ref{fig:rough}). 
This value corresponds to $1\%$ of the side of the simulation domain for $r=0.02$.
Indicatively, the side of the simulation domain can be assumed to be $350$ [$\mu$m],
leading to a depth of the defects of the surface of $3.5$ [$\mu$m]. In the case of
graphite-polymer composites, we observe that at the micron scale the surface of the fillers is 
irregular (see Fig.~\ref{fig:micro}).
\begin{table*}[b]
\begin{center}
\begin{tabular}{l|cc|c}
\hline\hline
 & Particle1 & Particle2 & Volume ratio \\
\hline
bimodal1 &  $(0.2,0.2,0.12)$ & $(0.06,0.06,0.04)$ & $33.3:1$  \\
\hline
bimodal2 & $(0.2,0.2,0.12)$ & $(0.1,0.1,0.06)$ & $8:1$  \\
\hline
bimodal3 & $(0.2,0.2,0.12)$ & $(0.14,0.14,0.08)$ & $3.1:1$ \\
\hline
granular & $(0.2,0.2,0.12)$ & $(0.2,0.2,0.12)$ & $1:1$ \\
\hline
fine & $(0.06,0.06,0.04)$ & $(0.06,0.06,0.04)$ &  $1:1$ \\
\hline\hline
\end{tabular}
\end{center}
\caption{\label{tab:powders}
Generated powders of different mixtures. The vectors $(l_{x},l_{y},l_{z})$ defines the average
side lengths of the blocks.}
\end{table*}
\begin{table*}[b]
\begin{center}
\begin{tabular}{l|l|ccc}
\hline\hline
No. & Type & $\theta_{x}$ & $\theta_{y}$ & $\theta_{z}$\\
\hline
$\mathrm{I}$ & aligned & $0$ & $0$ & $0$
\\
\hline
$\mathrm{II}$ & misaligned & $0$ & $0$ & $[-\pi/4,+\pi/4]$
\\
\hline
$\mathrm{III}$ & tilted $x$ & $[-\pi/36,+\pi/36]$ & $0$ & $0$
\\
\hline
$\mathrm{IV}$ & tilted $xy$ & $[-\pi/36,+\pi/36]$ & $[-\pi/36,+\pi/36]$ & $0$
\\
\hline
$\mathrm{V}$ & misaligned and tilted $x$ & $[-\pi/36,+\pi/36]$ & $0$ &$[-\pi/4,+\pi/4]$
\\
\hline
$\mathrm{VI}$ & misaligned and tilted $xy$ & $[-\pi/36,+\pi/36]$ & $[-\pi/36,+\pi/36]$ & $[-\pi/4,+\pi/4]$\\
\hline\hline
\end{tabular}
\end{center}
\caption{\label{tab:case}
Case studies and their orientation characteristics.}
\end{table*}
\begin{table*}[t]
\begin{center}
\begin{tabular}{l|cccccc}
\hline\hline
 & $\mathrm{I}$ & $\mathrm{II}$ & $\mathrm{III}$ & $\mathrm{IV}$ & $\mathrm{V}$ & $\mathrm{VI}$\\
\hline
bimodal1\hfill(std) & $55.1\%$ & $49.5\%$ & $54.4\%$ & $52.5\%$ & $50.1\%$ & $49.4\%$\\
\hline
bimodal1 (press) & $59.5\%$ & $54.3\%$ & $57.4\%$ & $57.2\%$  & $54.9\%$ & $52.1\%$\\
\hline
bimodal2\hfill(std) & $52.0\%$ & $49.3\%$ & $52.2\%$ & $51.0\%$ & $47.2\%$ & $47.9\%$ \\
\hline
bimodal2 (press) & $58.7\%$ & $51.4\%$ & $56.7\%$ & $55.7\%$  & $51.0\%$ & $50.3\%$\\
\hline
bimodal3\hfill(std) & $50.4\%$ & $45.0\%$ & $48.8\%$ & $48.8\%$ & $43.1\%$ & $42.9\%$\\
\hline
bimodal3 (press) & $54.9\%$ & $47.6\%$ & $52.9\%$ & $51.5\%$  & $46.1\%$ & $46.5\%$\\
\hline
granular\hfill(std) 
& $40.9\%$ & $37.0\%$ & $38.8\%$ & $41.5\%$ & $35.7\%$ & $34.7\%$\\
\hline
granular\hfill(press) 
& $44.5\%$ & $40.3\%$ & $43.0\%$ & $41.3\%$ & $42.4\%$ & $37.9\%$\\
\hline
fine\hfill(std) & $34.1\%$ & $33.4\%$ & $33.8\%$ & $33.8\%$ & $32.8\%$ & $33.3\%$\\
\hline
fine\hfill(press) & $36.6\%$ & $35.7\%$ & $36.4\%$ & $36.3\%$ & $35.8\%$ & $35.7\%$\\
\hline\hline
\end{tabular}
\end{center}
\caption{\label{tab:filling}
Packing fractions as obtained from simulations for systems with different orientational characteristics and size distributions.
The name of the mixtures is defined in Tab.~\ref{tab:powders}.
The labels $\mathrm{I}$-$\mathrm{VI}$ refer to the systems with the degree of orientational randomness defined
in Tab.~\ref{tab:case}. 
The abbreviation std stands for standard (without pressure and diffusive motion); press
means that an external pressure is applied as explained in Sec.~\ref{sec:settings}.}
\end{table*}
\begin{table*}[t]
\begin{ruledtabular}
\begin{center}
\begin{tabular}{ll|cc|cc|cc|cc|cc|cc}
& & \multicolumn{2}{c|}{$\mathrm{I}$} & \multicolumn{2}{c|}{$\mathrm{II}$} & \multicolumn{2}{c|}{$\mathrm{III}$} 
& \multicolumn{2}{c|}{$\mathrm{IV}$} & \multicolumn{2}{c|}{$\mathrm{V}$} & \multicolumn{2}{c}{$\mathrm{VI}$}\\
& & no.~part. & density & no.~part. & density & no.~part. & density & no.~part. & density & no.~part. & density & no.~part. & density\\
\hline
\multirow{2}*{bimodal1 (std)} & large & $101$   & $41.8\%$ & $81$    & $34.2\%$ & $93$    & $40.1\%$ & $91$    & $37.9\%$ & $81$    & $35.0\%$ & $77$    & $34.2\%$\\
                              & small & $1'222$ & $13.3\%$ & $1'423$ & $15.3\%$ & $1'325$ & $14.3\%$ & $1'345$ & $14.6\%$ & $1'436$ & $15.1\%$ & $1'448$ & $15.2\%$\\
\hline
\multirow{2}*{bimodal1 (press)} & large & $110$    & $45.4\%$ & $88$    & $38.9\%$ & $100$   & $41.8\%$ & $102$   & $42.9\%$ & $92$    & $41.1\%$ & $81$    & $35.9\%$\\
                                & small & $1'263$  & $14.1\%$ & $1'405$ & $15.4\%$ & $1'415$ & $15.6\%$ & $1'290$ & $14.3\%$ & $1'271$ & $13.8\%$ & $1'498$ & $16.2\%$\\
\hline
\multirow{2}*{bimodal2 (std)} & large & $93$  & $38.1\%$ & $85$  & $36.7\%$ & $90$  & $37.7\%$ & $90$  & $37.4\%$ & $81$  & $34.6\%$ & $83$  & $36.0\%$\\
                              & small & $346$ & $13.9\%$ & $310$ & $12.6\%$ & $368$ & $14.5\%$ & $335$ & $13.6\%$ & $288$ & $12.6\%$ & $292$ & $11.9\%$\\
\hline
\multirow{2}*{bimodal2 (press)} & large & $110$ & $45.7\%$ & $88$  & $38.5\%$ & $103$ & $43.4\%$ & $100$ & $43.2\%$ & $91$  & $40.0\%$ & $88$  & $38.9\%$\\
                                & small & $325$ & $13.0\%$ & $310$ & $12.9\%$ & $334$ & $13.3\%$ & $299$ & $12.5\%$ & $265$ & $11.0\%$ & $275$ & $11.4\%$\\
\hline
\multirow{2}*{bimodal3 (std)} & large & $98$ & $41.1\%$ & $85$ & $36.9\%$ & $88$ & $37.7\%$ & $89$ & $37.9\%$ & $81$ & $35.9\%$ & $76$ & $33.7\%$\\
                              & small & $80$ & $9.3\%$  & $65$ & $8.1\%$  & $96$ & $11.1\%$ & $93$ & $10.9\%$ & $57$ & $7.2\%$  & $73$ & $9.2\%$\\
\hline
\multirow{2}*{bimodal3 (press)} & large & $104$ & $43.5\%$ & $89$ & $38.6\%$ & $98$ & $42.1\%$ & $103$ & $44.1\%$ & $92$ & $40.2\%$ & $84$ & $38.3\%$\\
                                & small & $98$  & $11.4\%$ & $71$ & $9.0\%$  & $92$ & $10.8\%$ & $61$  & $7.4\%$  & $47$ & $5.9\%$  & $66$ & $8.2\%$\\
\hline
\multirow{2}*{granular (std)} 
& large & $99$ & $40.9\%$ & $88$ & $37.0\%$ & $92$ & $38.8\%$ & $98$ & $41.5\%$ & $81$ & $35.7\%$ & $80$ & $34.7\%$\\
& small &  $0$ & $0\%$    &  $0$ & $0\%$    &  $0$ & $0\%$    & $0$  & $0\%$    & $0$  & $0\%$    & $0$  & $0\%$\\
\hline
\multirow{2}*{granular (press)} 
& large & $107$ & $44.5\%$ & $93$ & $40.3\%$ & $104$ & $43.0\%$ & $98$ & $41.3\%$ & $90$ & $42.4\%$ & $86$ & $37.9\%$\\
& small & $0$   & $0\%$    & $0$  & $0\%$    & $0$   & $0\%$    & $0$  & $0\%$    & $0$  & $0\%$    & $0$  & $0\%$    \\
\hline
\multirow{2}*{fine (std)} & large & $0$     & $0\%$    & $0$     & $0\%$    & $0$      & $0\%$    & $0$     & $0\%$    & $0$     & $0\%$    & $0$     & $0\%$\\
                          & small & $3'066$ & $34.1\%$ & $3'000$ & $33.4\%$ & $3'014$  & $33.8\%$ & $3'063$ & $33.8\%$ & $2'962$ & $32.8\%$ & $2'993$ & $33.3\%$\\
\hline
\multirow{2}*{fine (press)} & large & $0$     & $0\%$    & $0$     & $0\%$    & $0$     & $0\%$    & $0$     & $0\%$    & $0$     & $0\%$    & $0$     & \\
                            & small & $3'231$ & $36.6\%$ & $3'148$ & $35.7\%$ & $3'182$ & $36.4\%$ & $3'191$ & $36.3\%$ & $3'144$ & $35.8\%$ & $3'138$ & $35.7\%$\\
\end{tabular}
\end{center}
\caption{\label{tab:density}
Number of block particles of type larger and smaller composing every bimodal powder and their contribution to the packing density.
Misalignment of larger blocks is the primary cause of significant variations of the overall volume fraction. The mixtures
and the size of their particles are introduced in Tab.~\ref{tab:powders}. Roman numbers specify the degree of orientational 
randomness as defined in Tab.~\ref{tab:case}.}
\end{ruledtabular}
\end{table*}
\begin{figure*}[b]
\includegraphics[width=4cm]{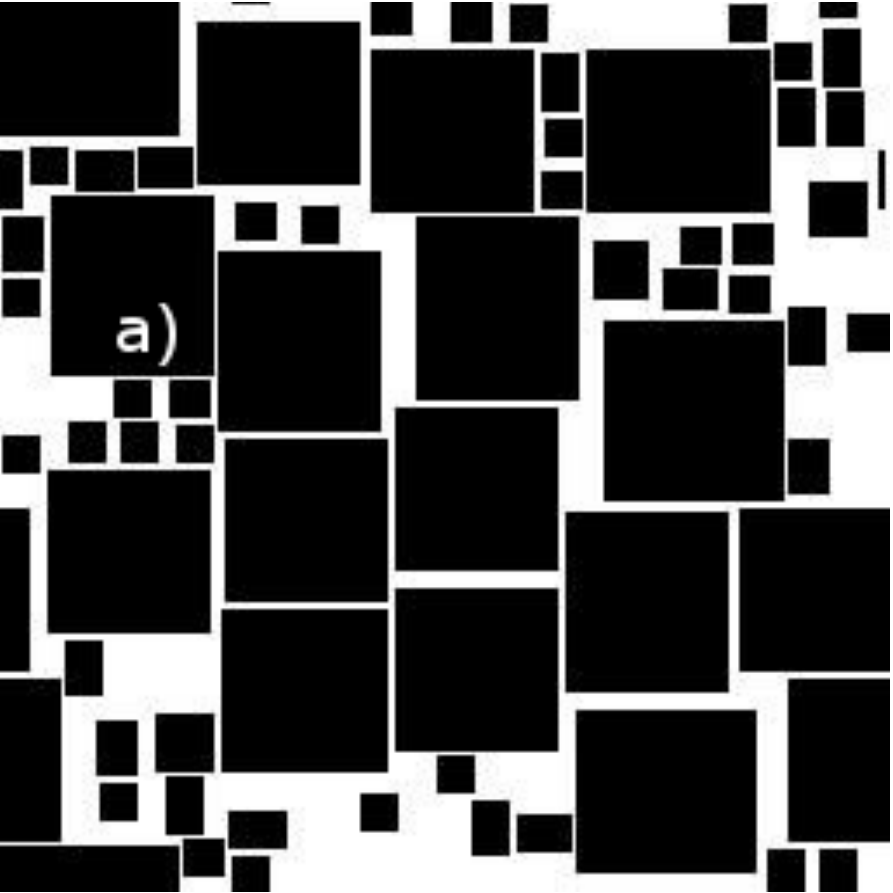}\hspace{0.5cm}
\includegraphics[width=4cm]{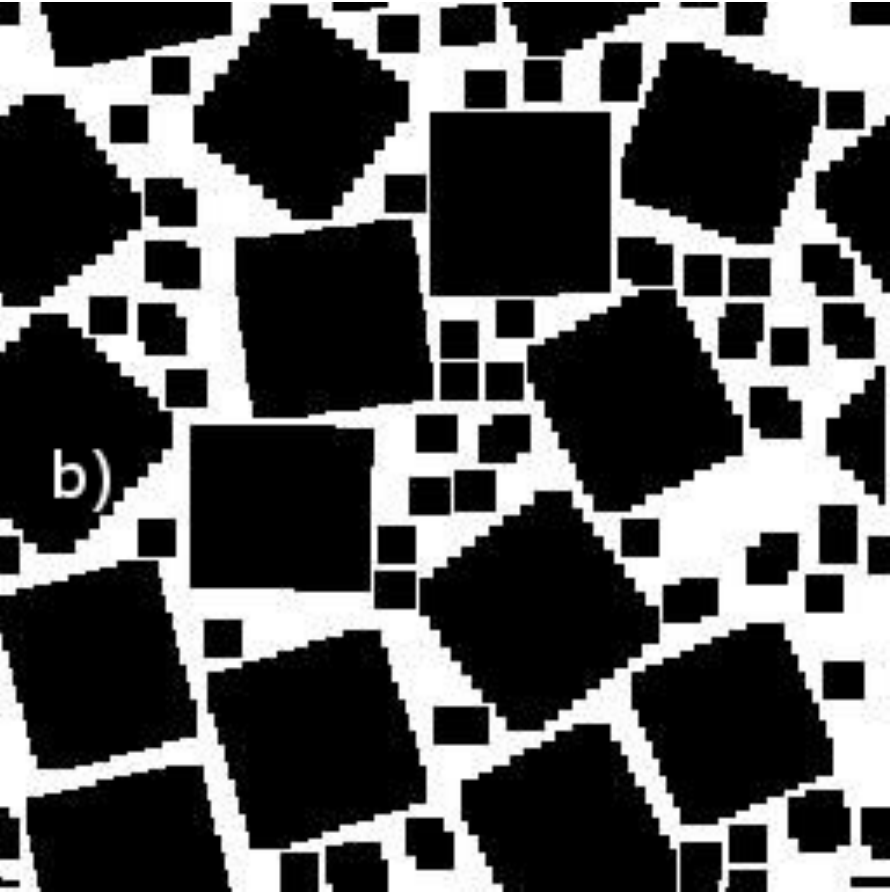}\hspace{0.5cm}
\includegraphics[width=4cm]{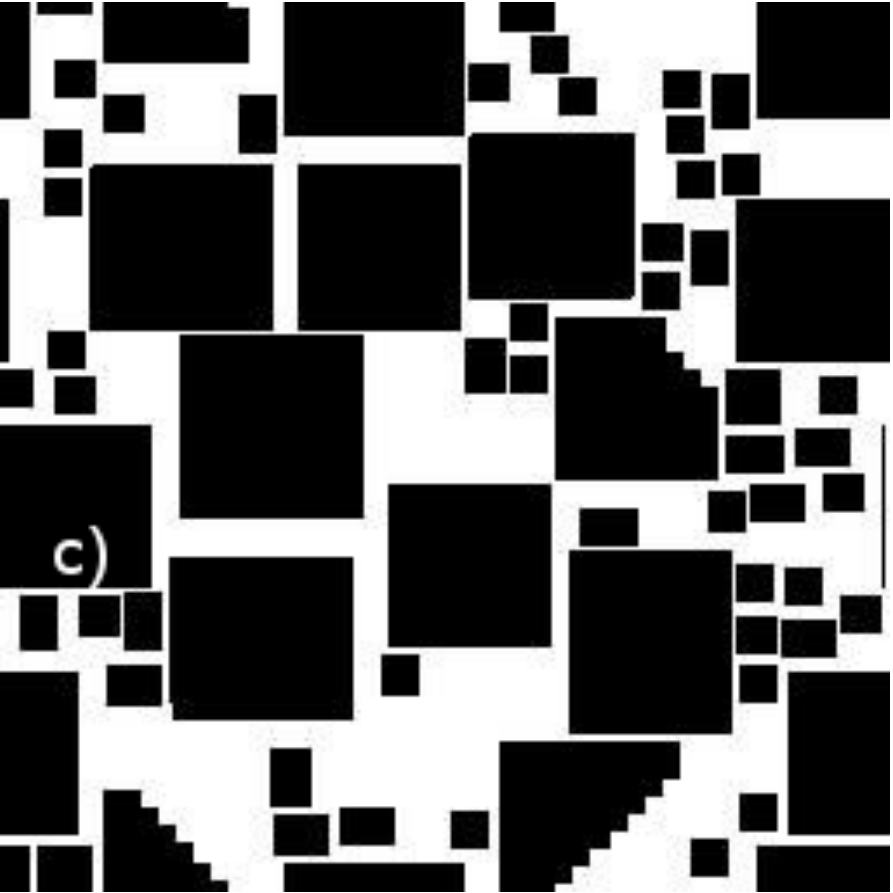}\hspace{0.5cm}
\includegraphics[width=4cm]{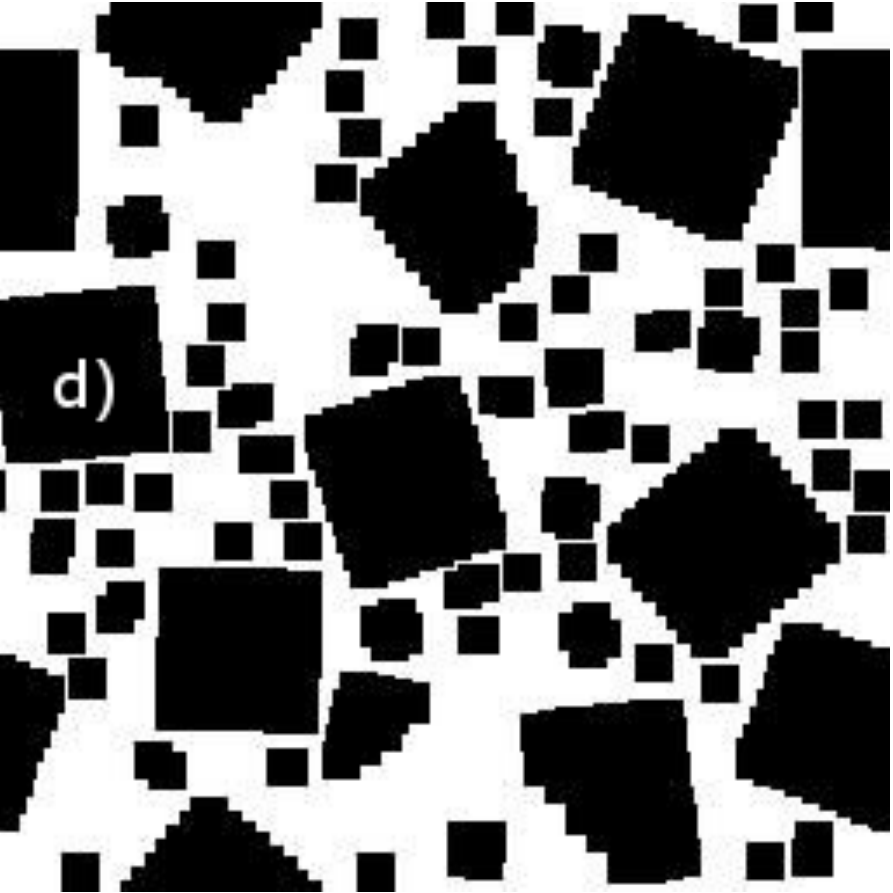}\\
\caption{Sections of some packing structures for the mixture bimodal1 
(see Tab.~\ref{tab:powders}) with an applied external pressure.
a) Aligned case. b) Misaligned case. c) Tilted $xy$ case. Note that
the boundary of the blocks in the plane $xy$ can be diagonal.
d) Misaligned and tilted $xy$ case. Note that the boundary of the blocks
can be irregular in this case.
\label{fig:layers}}
\end{figure*}

Diffusive motion is introduced by means of random moves of the blocks along the
$x$, $y$ and $z$ directions. The effect of an external pressure is accounted for by
introducing a bias toward the center of the domain. More precisely, after a 
given number of unsuccessful attempts, all the already placed blocks are moved
randomly. For every block, first the direction of the trial move is chosen randomly.
The sense of the move is then chosen with probabilities $p_{x}$, $p_{y}$ and $p_{z}$.
For example, for a block with the first sphere in the region defined by $x>1/2$,
$p_{x}$ is the probability to be translated toward the center with sense and
direction defined by the vector $(-1,0,0)$. Given a block, the distance spanned by the 
trial moves is different for every direction. When a move is not possible because 
resulting in overlap, the distance associated with that direction is reduced by 
a factor of $2$. The first time one of the possible move distances falls below 
$d_{\mathrm{min}}$, the block is no more taken into account for random moves. Finally, 
in this way, purely diffusive motion corresponds to $p_{x}=p_{y}=p_{z}=0.5$ while 
in all other cases there appears a resultant pressure (inward or outward oriented).

\section{Simulation settings and modeling approach}\label{sec:settings}

In all simulations reported here, the standard deviation $\sigma$ is set to the value $\sqrt{0.3}$ and the 
radius of the spheres is $r=0.02$. 
We always start with a given average size and smaller blocks are considered when no block can be placed after
$200'000$ trials. The simulation is terminated when no smaller block can be placed for
the same condition. Table \ref{tab:powders} lists the powders with the average size of blocks they are composed of.
Mixtures and aspect ratios are chosen so as to obtain composite materials produced experimentally 
by using common graphitic powders.  

Different random orientations are investigated. We distinguish the types of
arrangement listed in Tab.~\ref{tab:case}.
Every system is considered without diffusion and pressure and with pressure. 
In the latter case, for the probabilities of rearrangement we choose $p_{x}=p_{y}=p_{z}=1$ in order to reach 
the maximal packing fractions.
The initial possible displacement along the three directions is $0.5$. The minimal distance of approach is set to
$d_{\mathrm{min}}=2/N$, with $N=256$, corresponding to the width of two adjacent voxels. The blocks are
never moved twice consecutively, but one unsuccessful attempt is sufficient in order
to call the routine performing diffusive motion. The aim is to keep the pressure effect
as close as possible to the linear dependence on the total number of blocks when present 
in relatively large number.

Of course, the application of an external load has the effect to deform the fillers.
Furthermore, anisotropic pressures can also lead to a certain degree of orientational order.
We apply an isotropic pressure, but the shape of the fillers and the restricted
range of the tilt angles intend to simulate the effect of uniaxial compaction along the $z$
direction. Graphitic powders are the result of an extremely expensive process of
milling. Because of their molecular structure, the size of particles is reduced through both
exfoliation and fragmentation. It is clear that the combination of these phenomena introduces
faces and angularities in the resulting particles. In that respect, the block is the simplest
geometry approaching the shape of real graphitic assemblies.

\section{Results and discussion}

\begin{figure}[t]
\includegraphics[width=7cm]{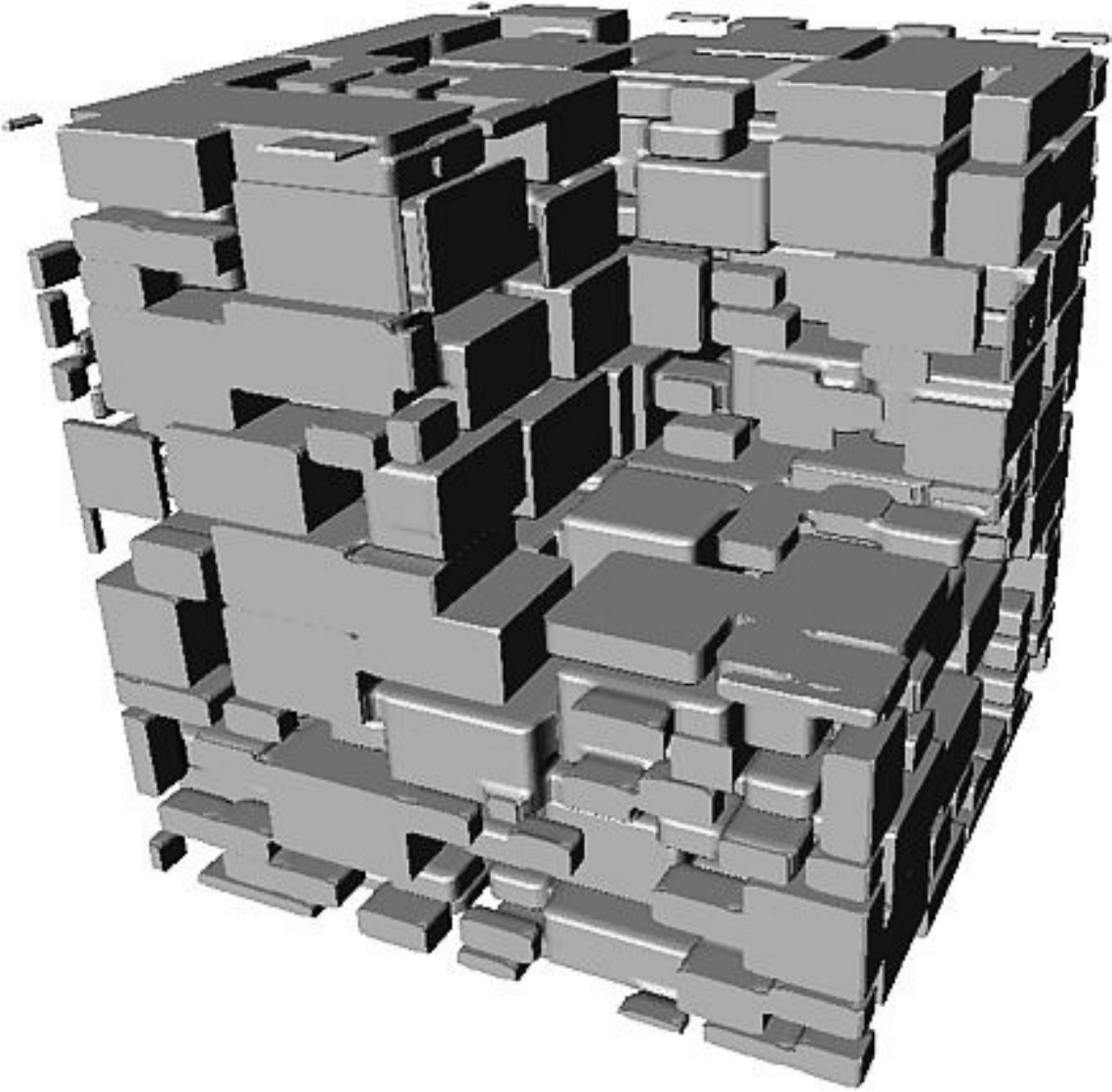}
\caption{$3$D rendering of the final structure for the powder bimodal2 with pressure in the
aligned case. The volume fraction is $58.7\%$.
\label{fig:3d}}
\end{figure}
\begin{figure}[t]
\includegraphics[width=8.5cm]{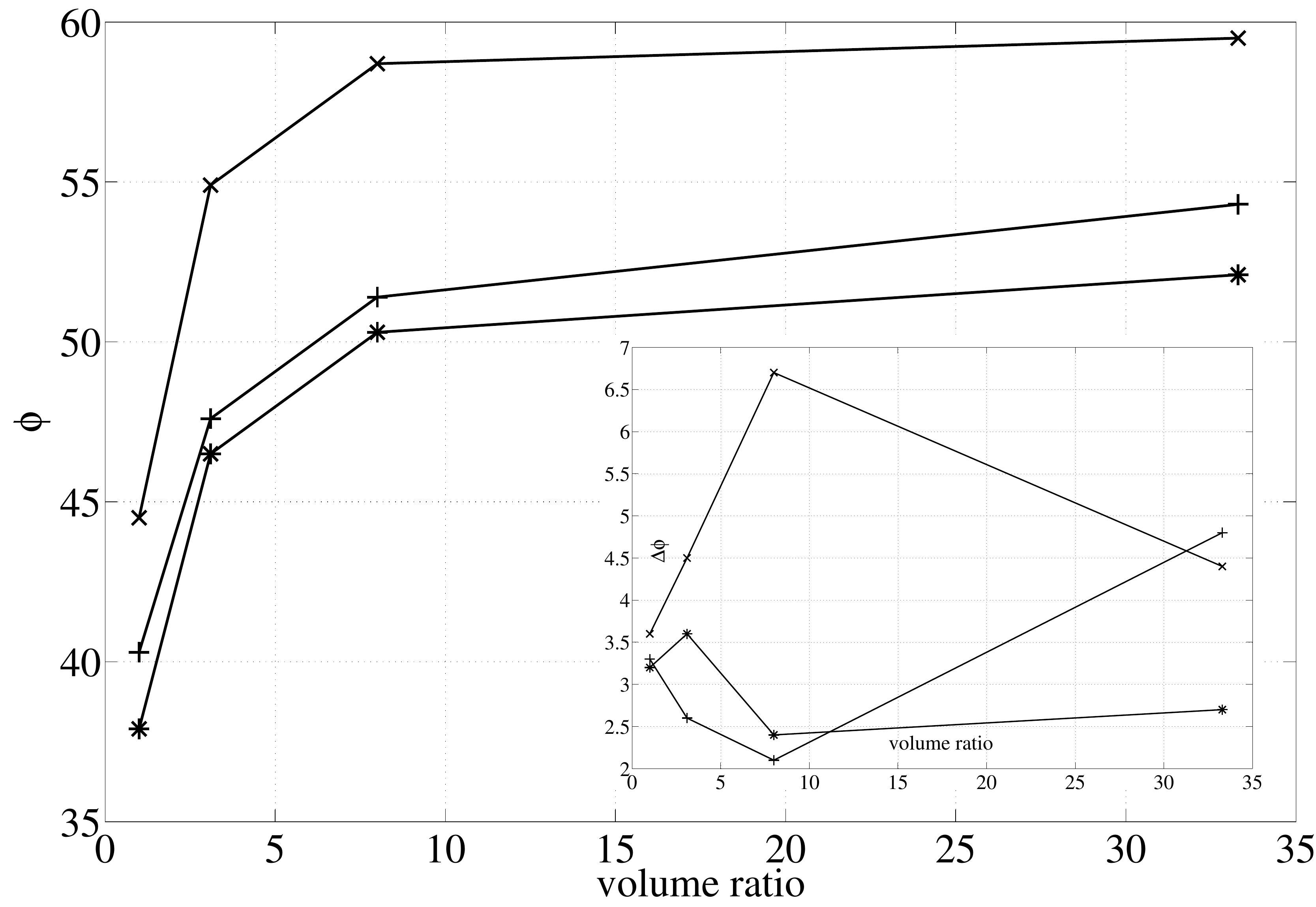}
\caption{Packing fraction against the volume ratio; pressure is applied. Aligned case ($\times$),
misaligned case ($+$) and misaligned with tilt xy case ($\ast$) (cf.~Fig.~\ref{fig:layers}). 
Inset: $\Delta\phi=\phi_{\mathrm{press}}-\phi_{\mathrm{std}}$ as a function of the volume
ratio. $\phi_{\mathrm{press}}$ is the packing fraction reached with pressure and 
$\phi_{\mathrm{std}}$ that realized with no pressure.
\label{fig:v2phi}}
\end{figure}
\begin{figure}[t]
\includegraphics[width=8.5cm]{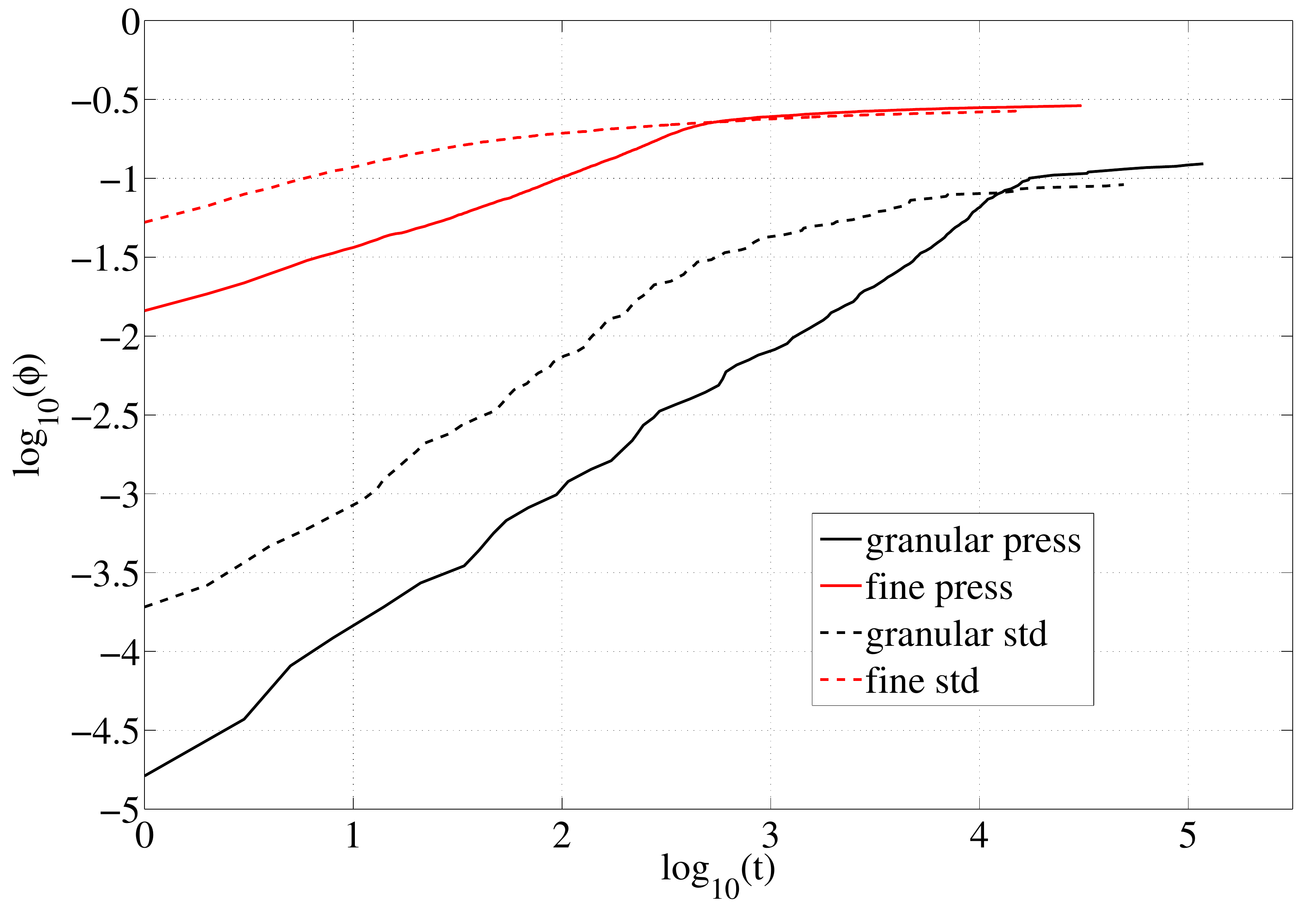}
\caption{ Packing fraction $\phi$ as a function of time
($\log$-$\log$ scale). All powders are of type
aligned (cf.~Tab.~\ref{tab:case}) with pressure (press) and with no pressure (std).
\label{fig:t2phi}}
\end{figure}
\begin{figure*}[ht]
\includegraphics[width=5.5cm]{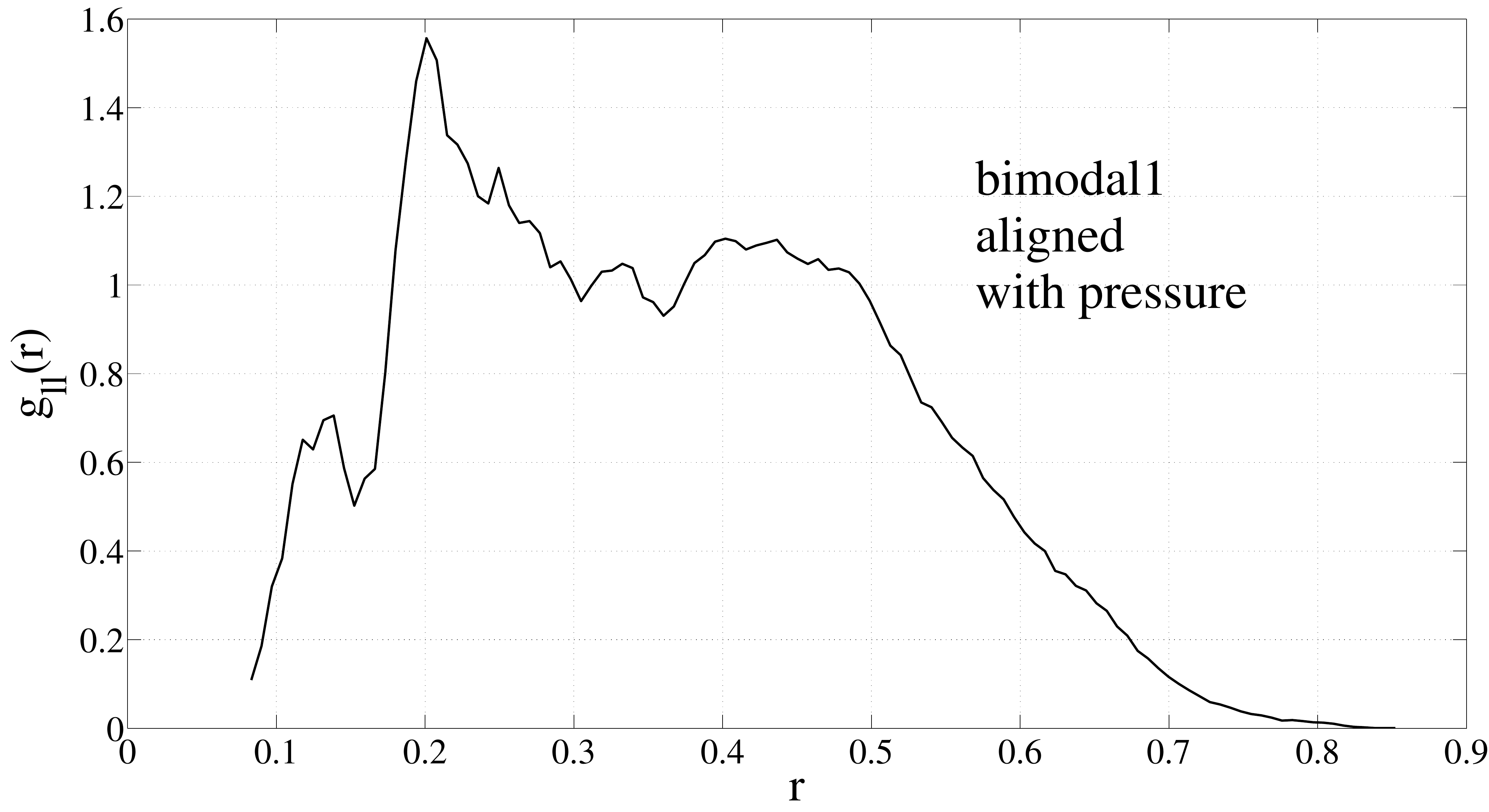}
\includegraphics[width=5.5cm]{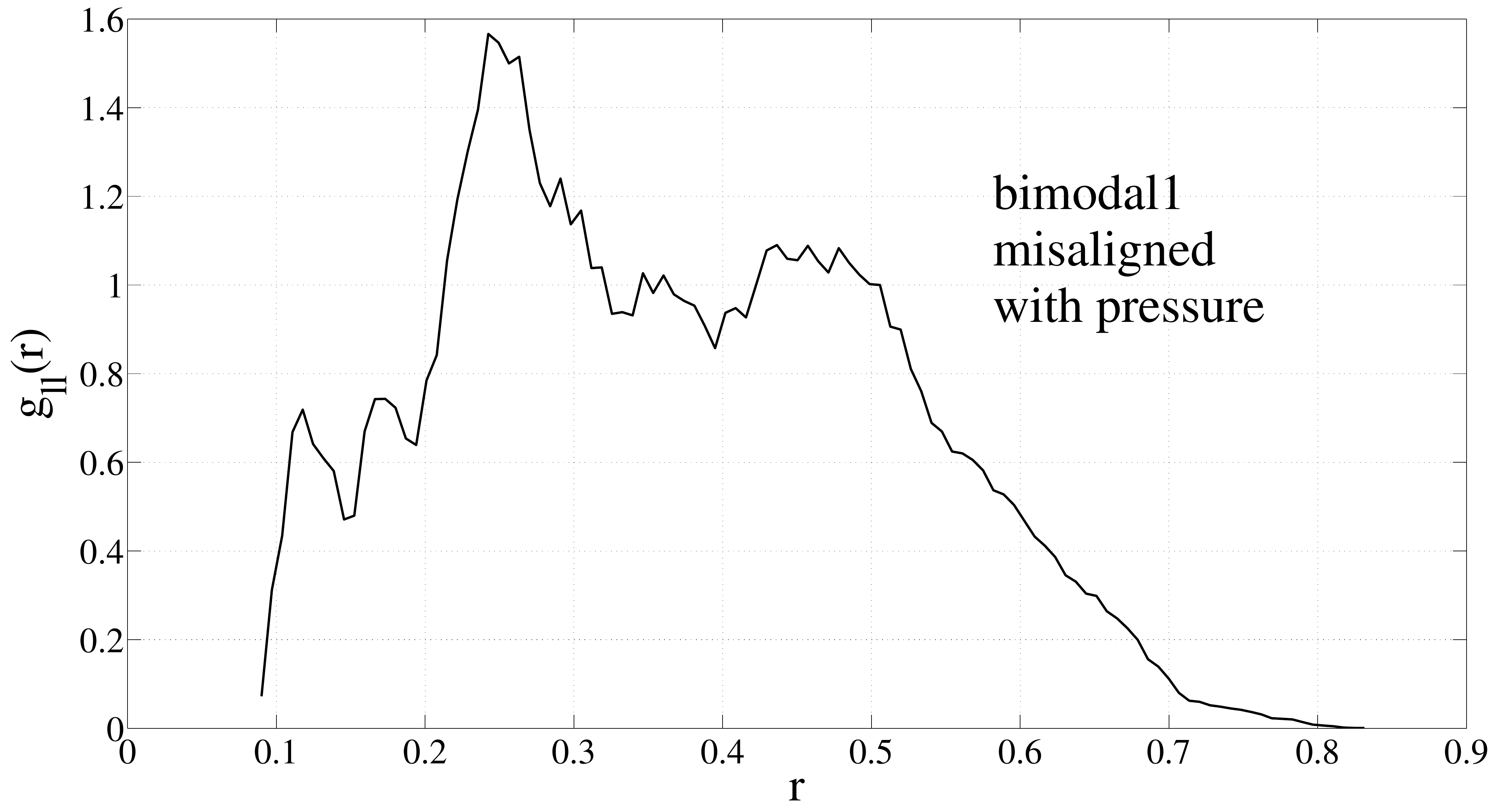}
\includegraphics[width=5.5cm]{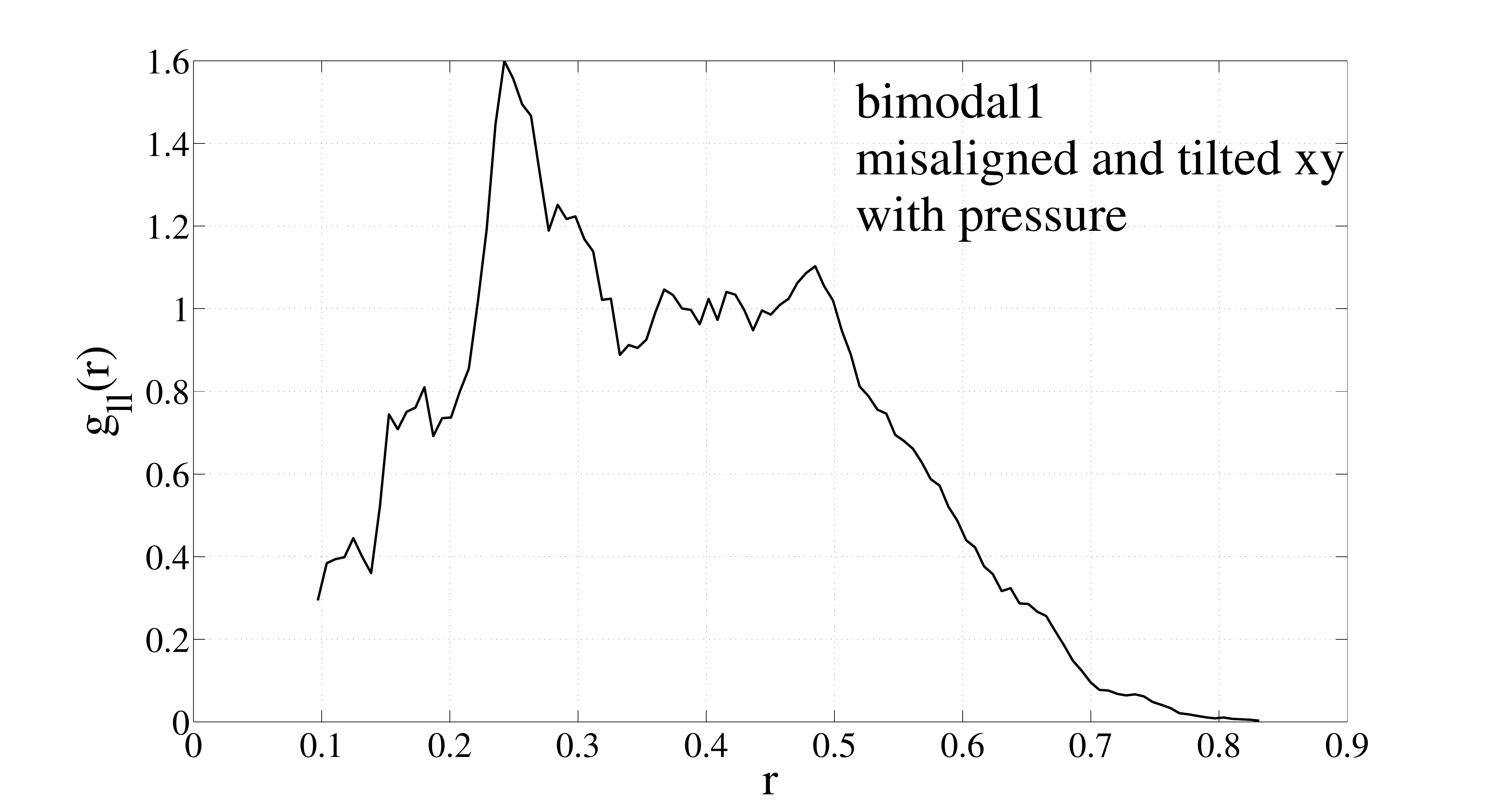}\\
\includegraphics[width=5.5cm]{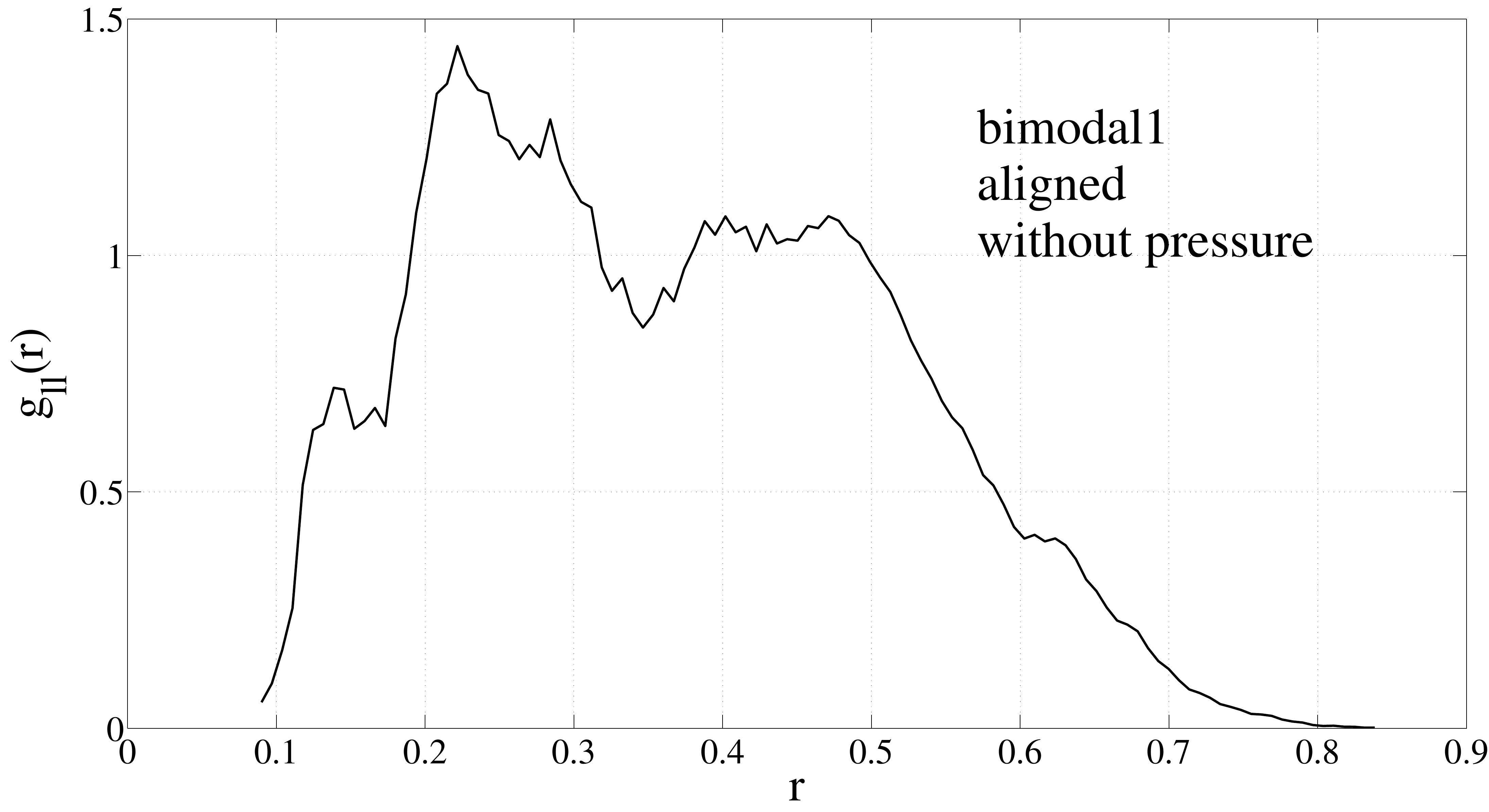}
\includegraphics[width=5.5cm]{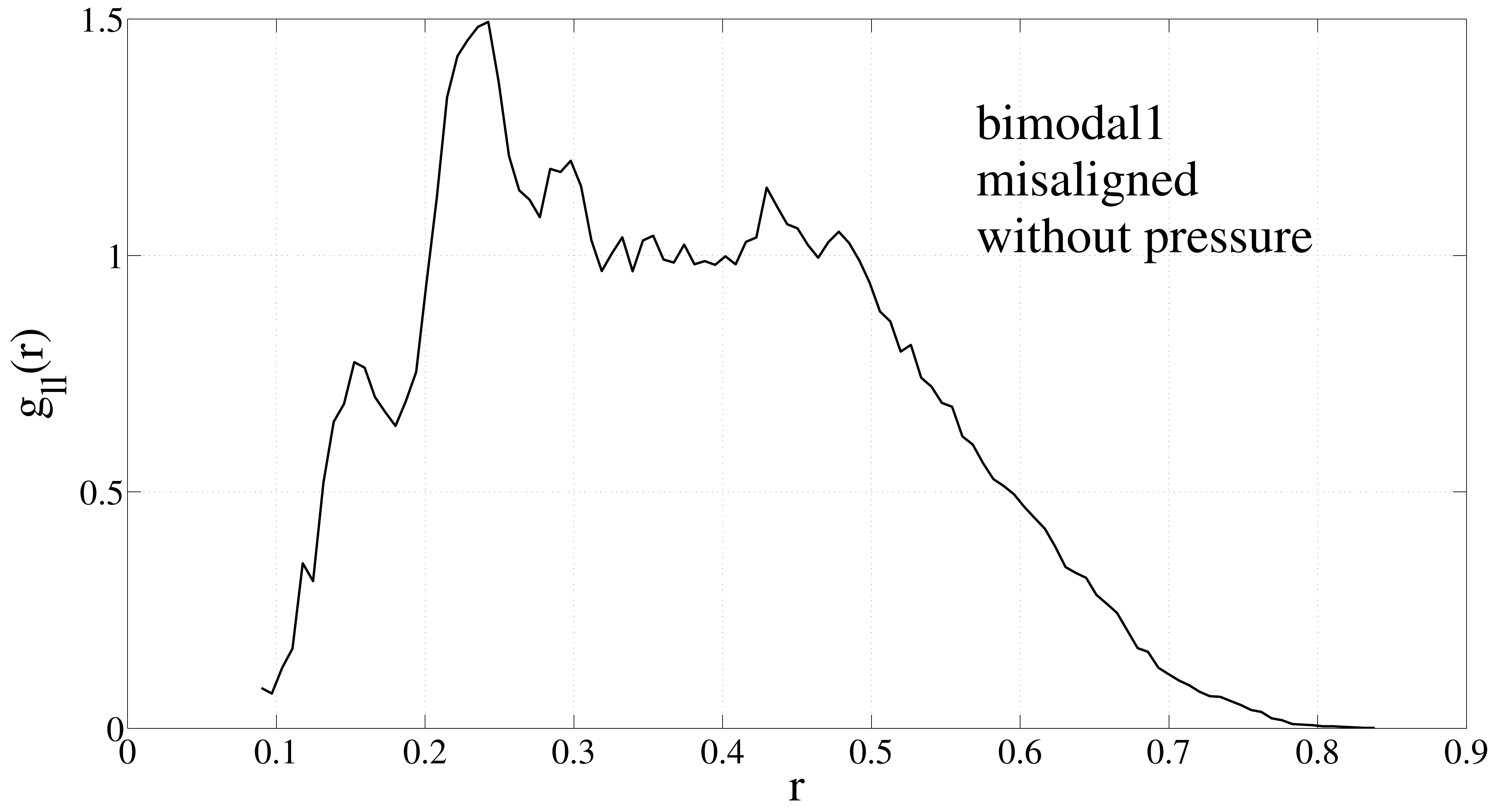}
\includegraphics[width=5.5cm]{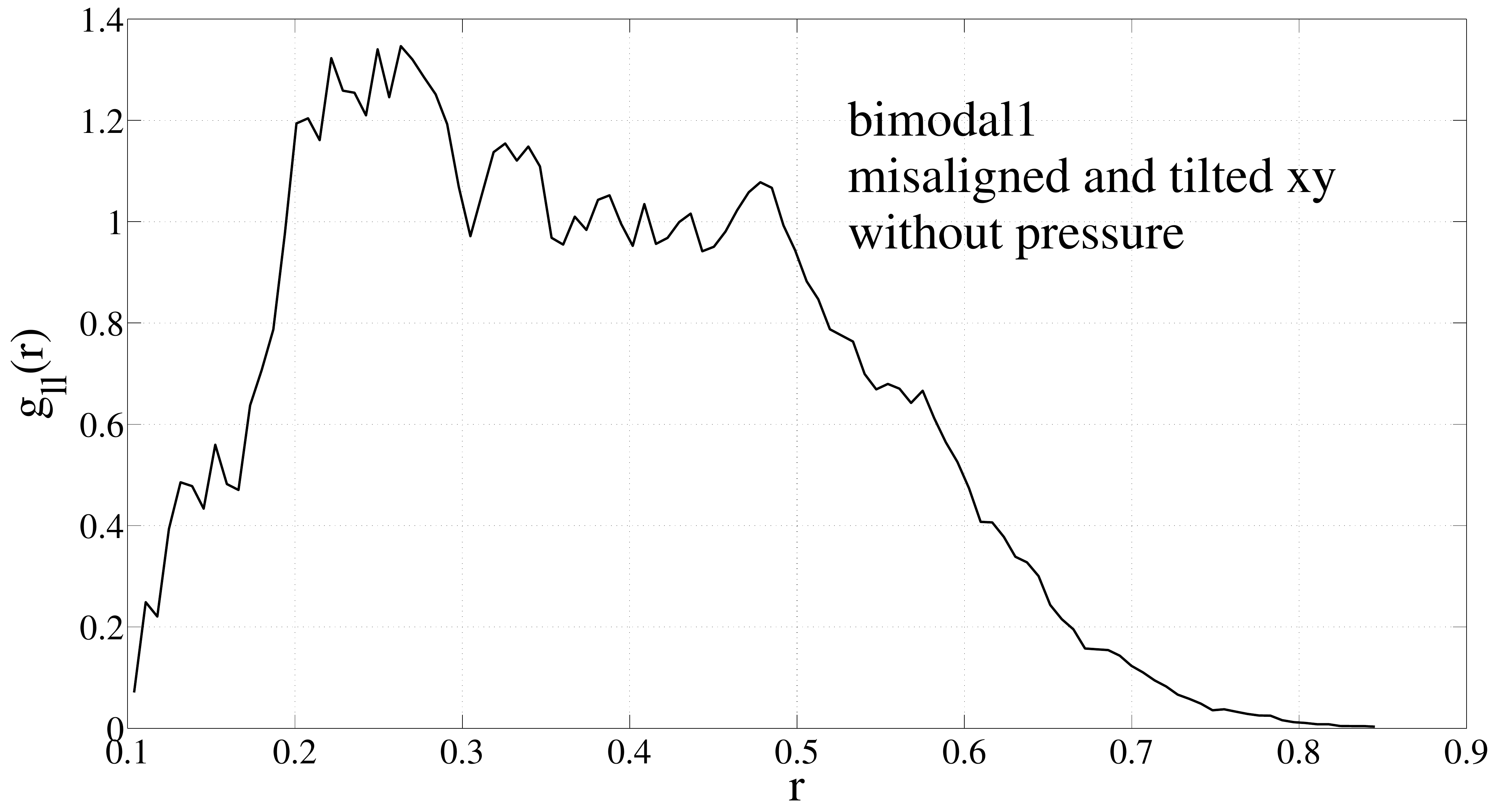}\\
\includegraphics[width=5.5cm]{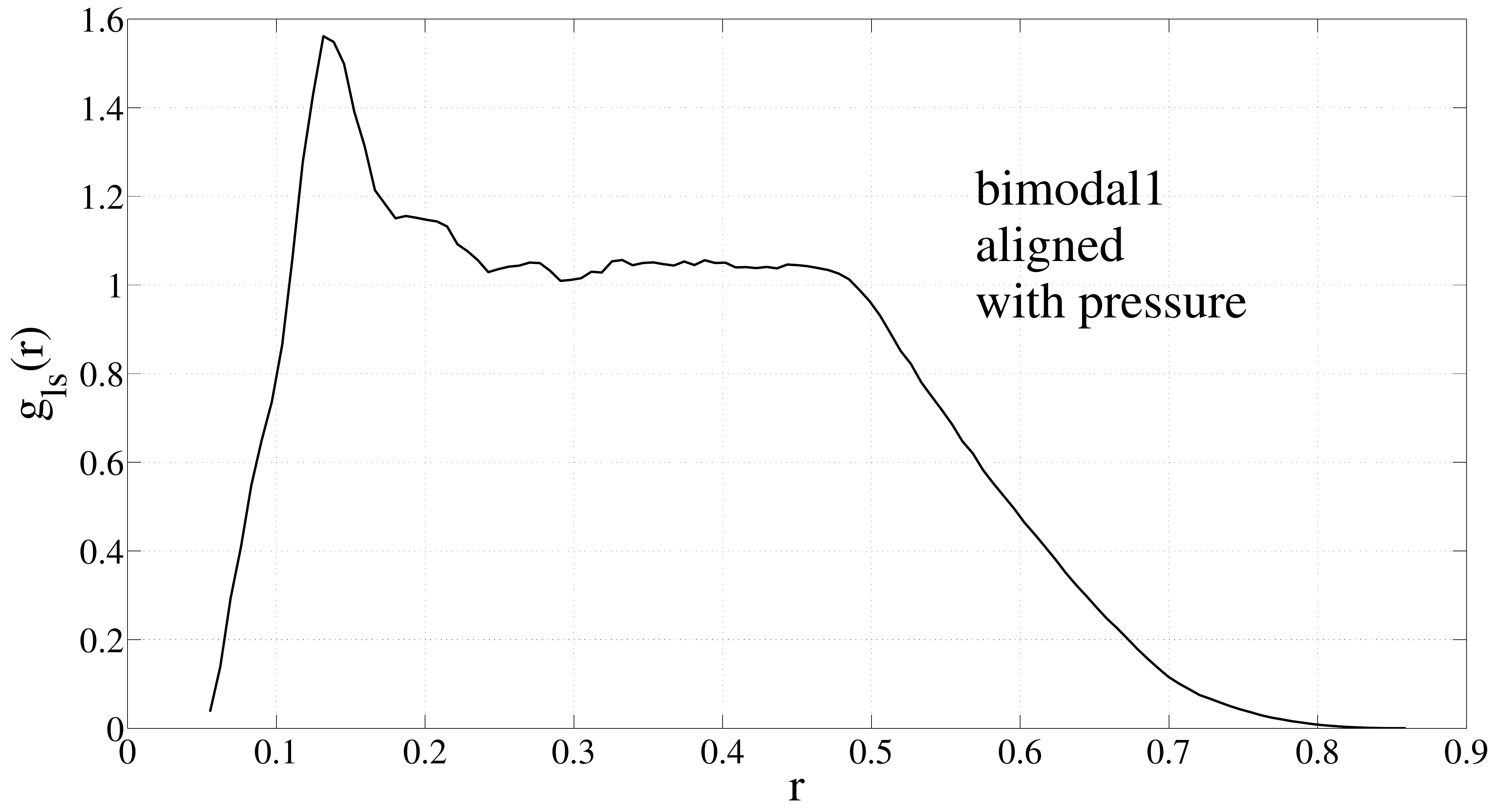}
\includegraphics[width=5.5cm]{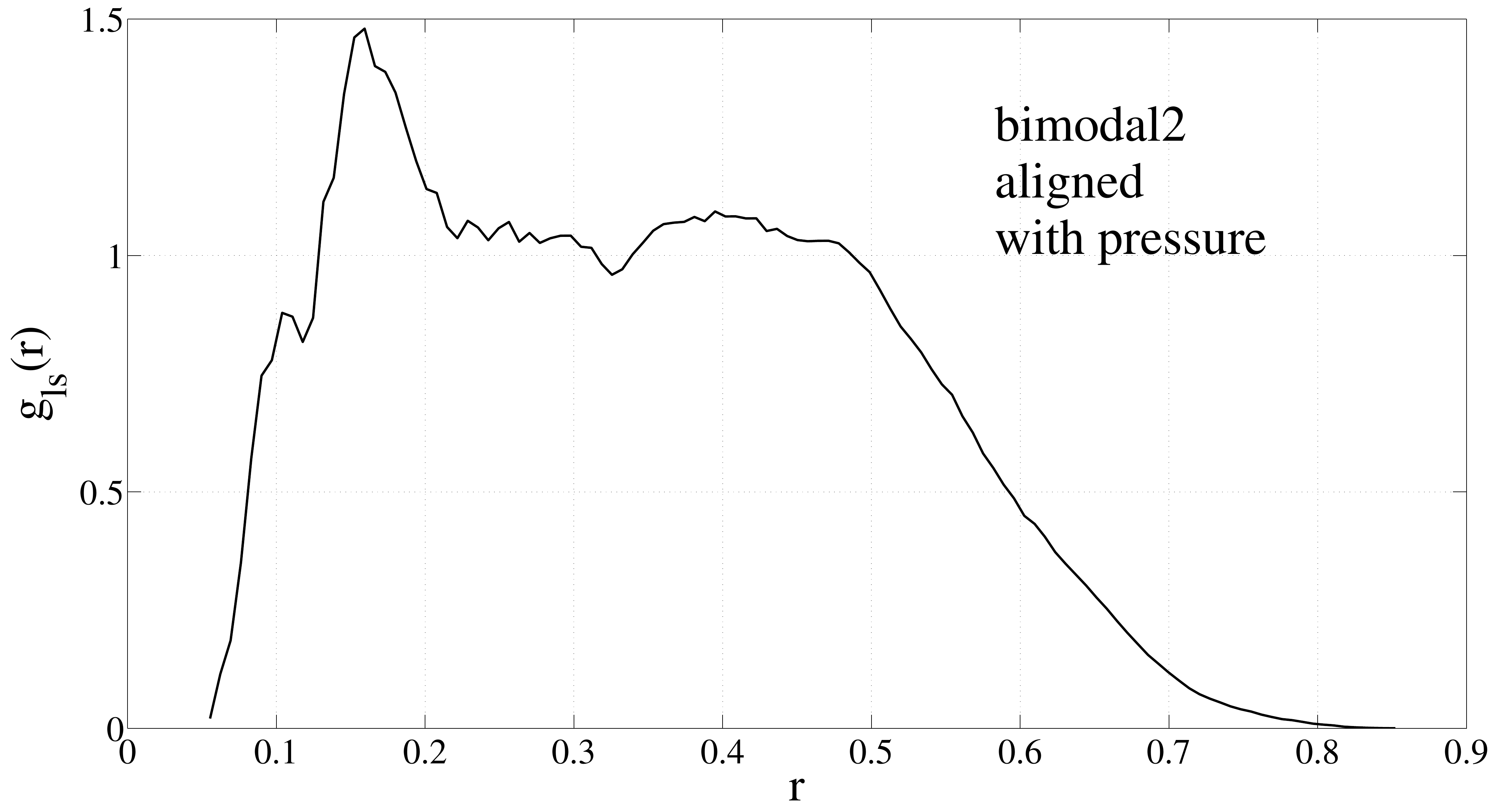}
\includegraphics[width=5.5cm]{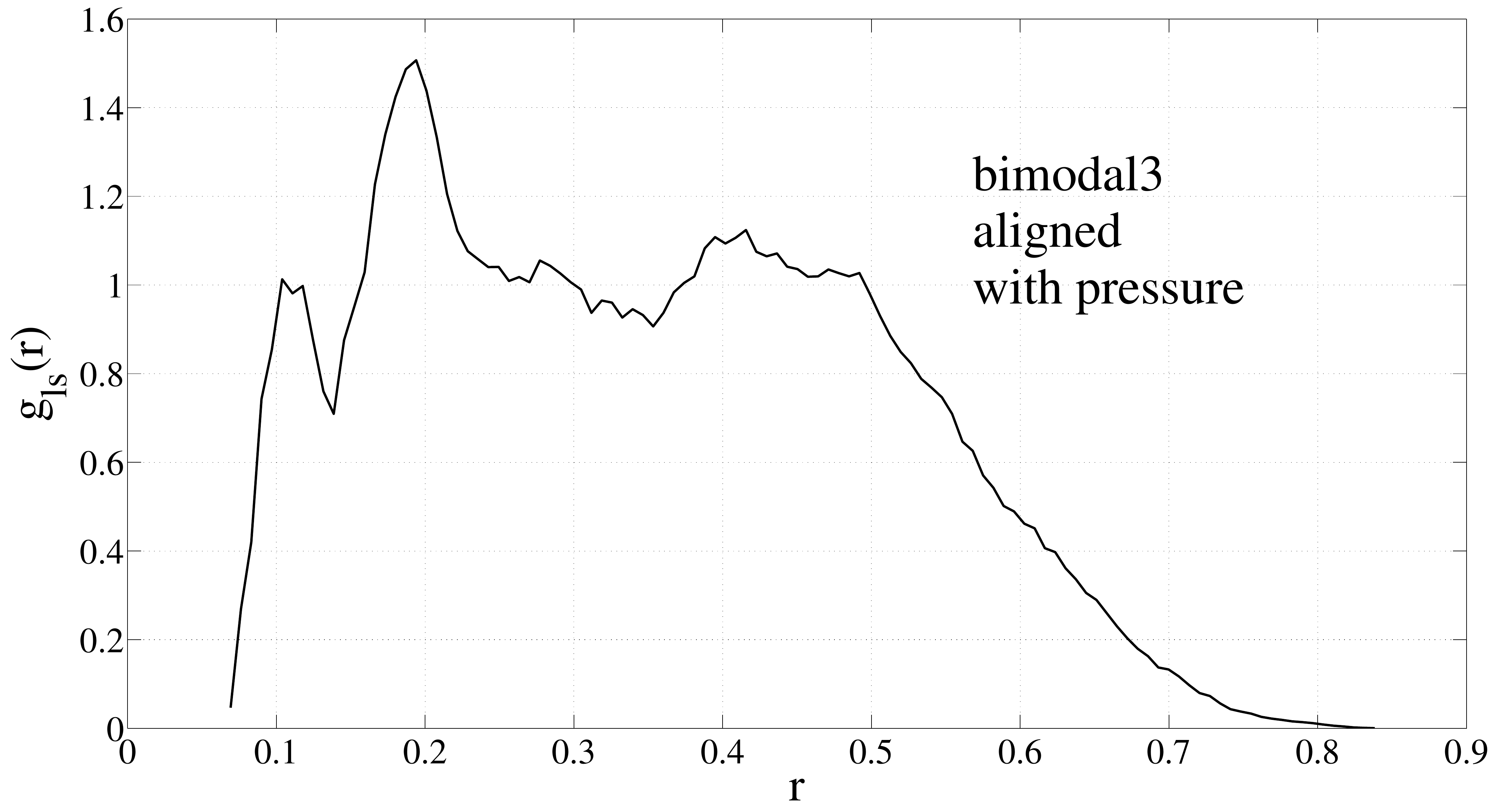}\\
\includegraphics[width=5.5cm]{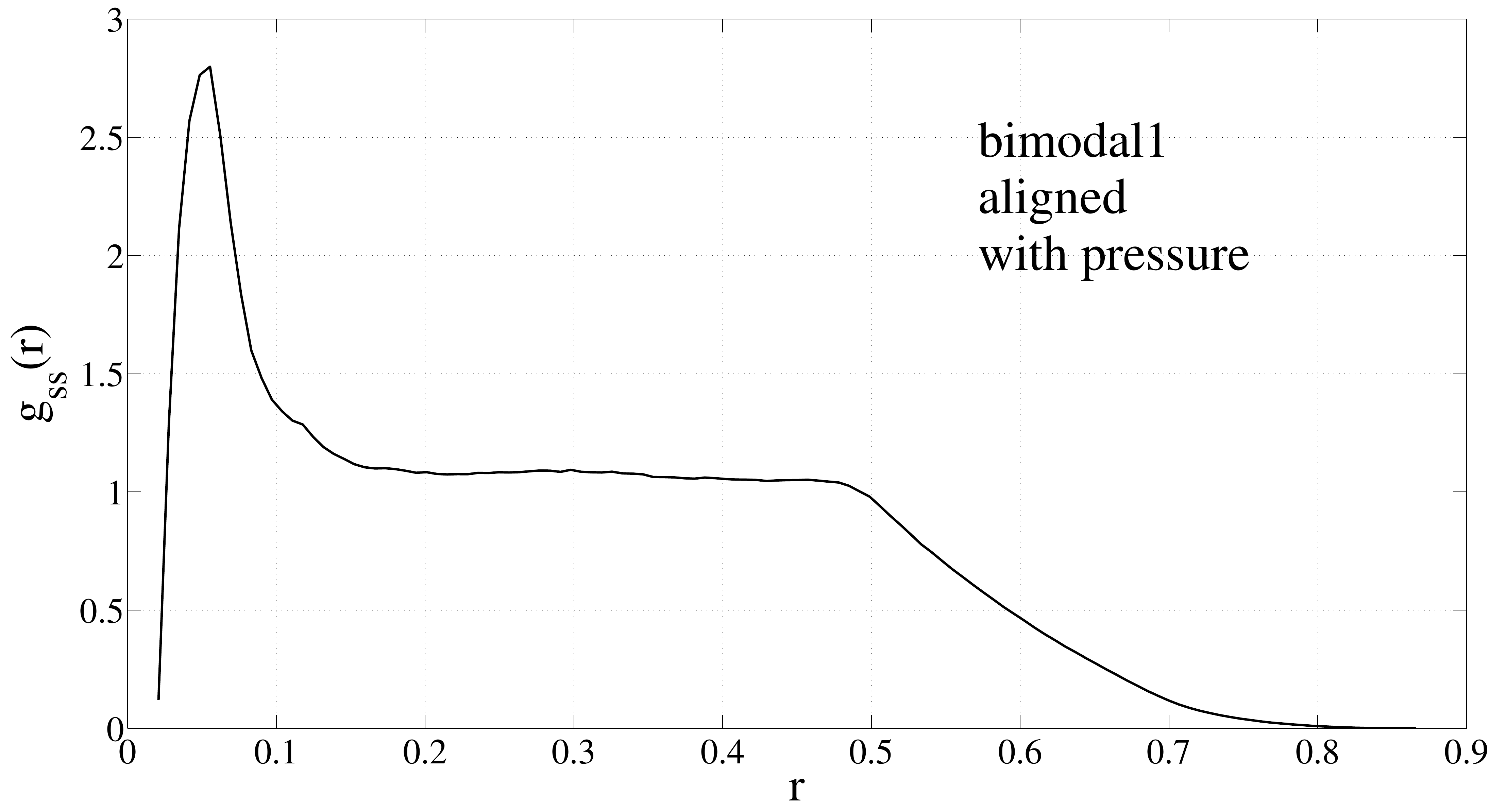}
\includegraphics[width=5.5cm]{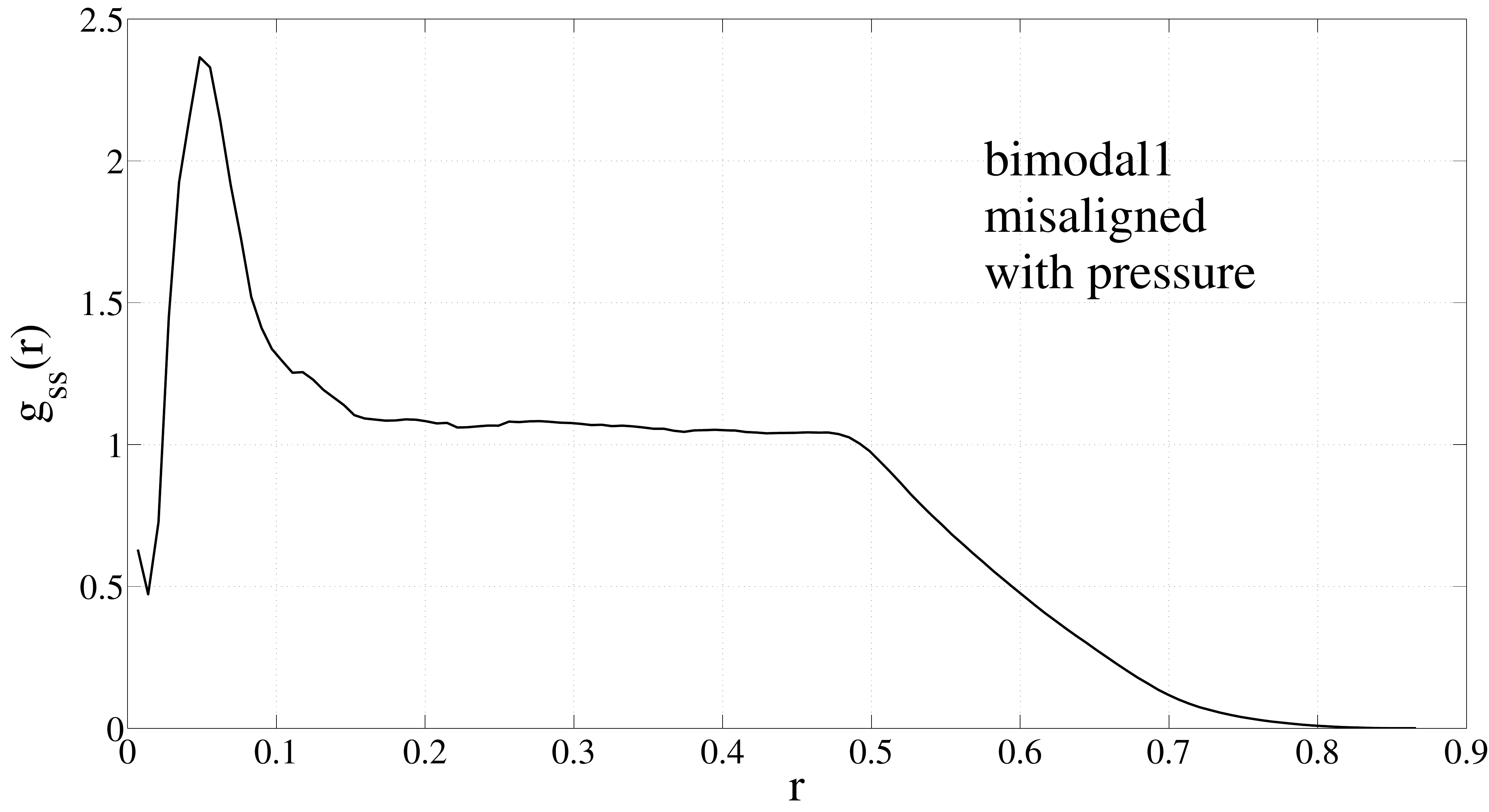}
\includegraphics[width=5.5cm]{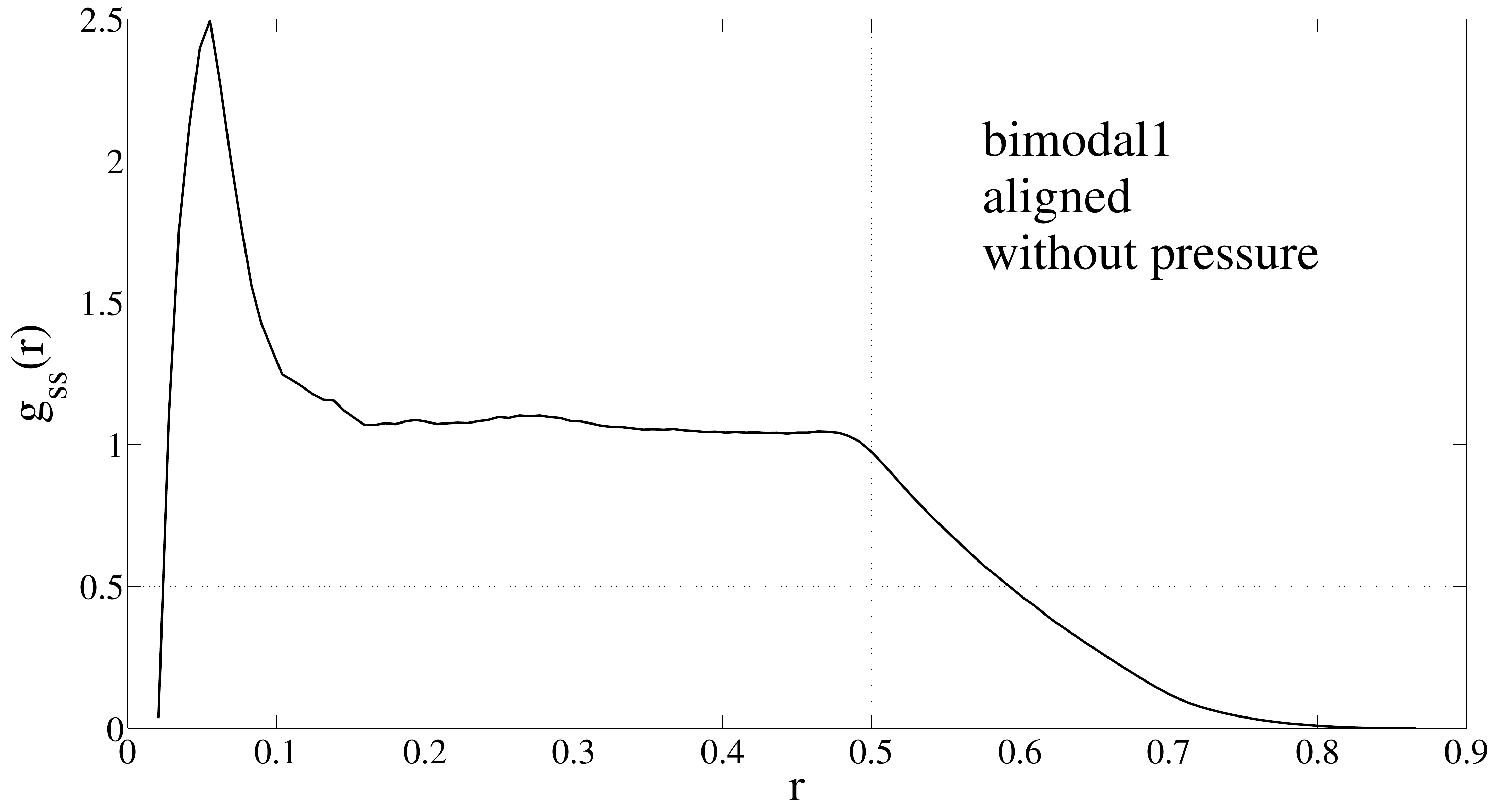}\\
\includegraphics[width=5.5cm]{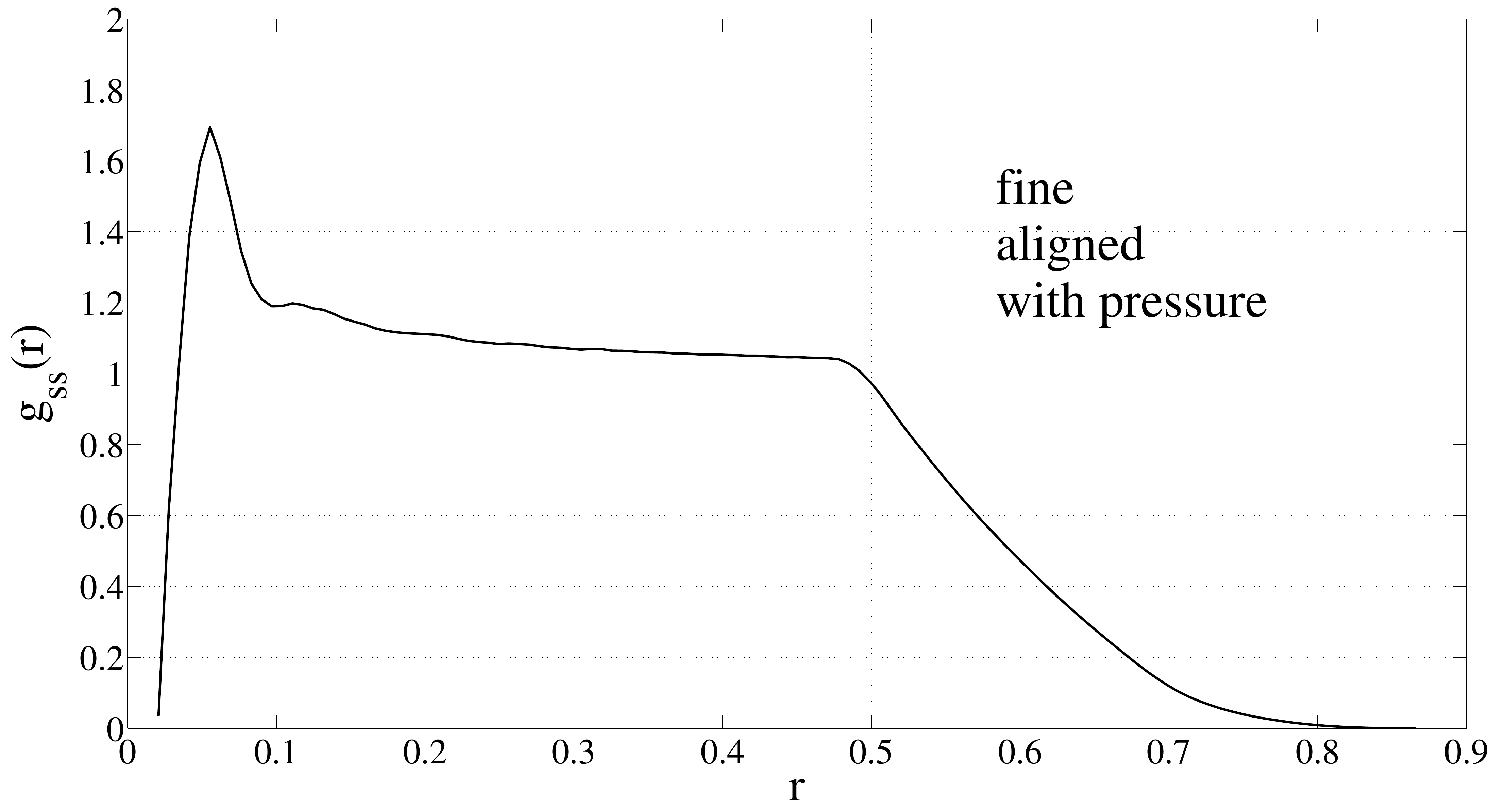}
\includegraphics[width=5.5cm]{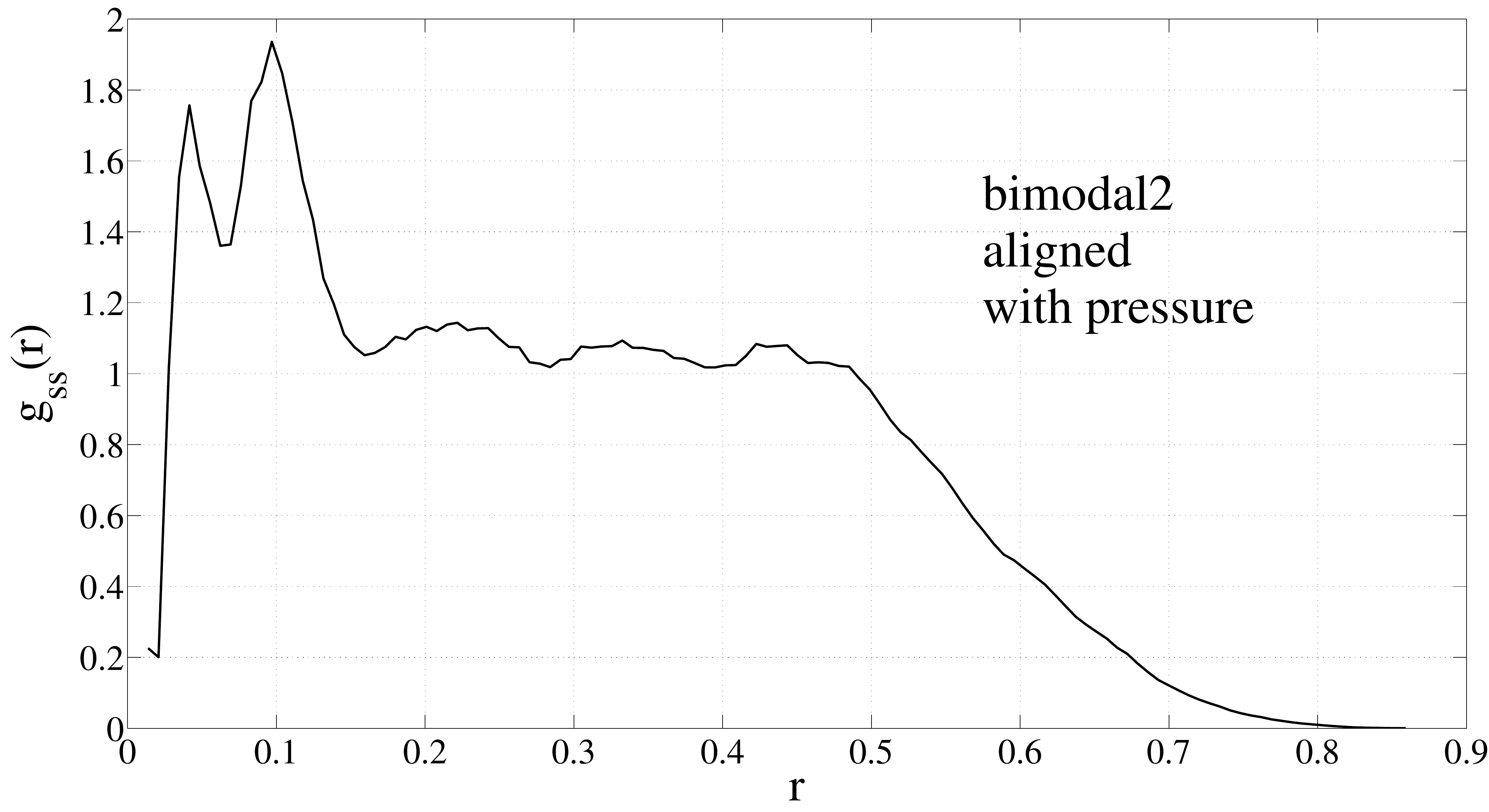}
\includegraphics[width=5.5cm]{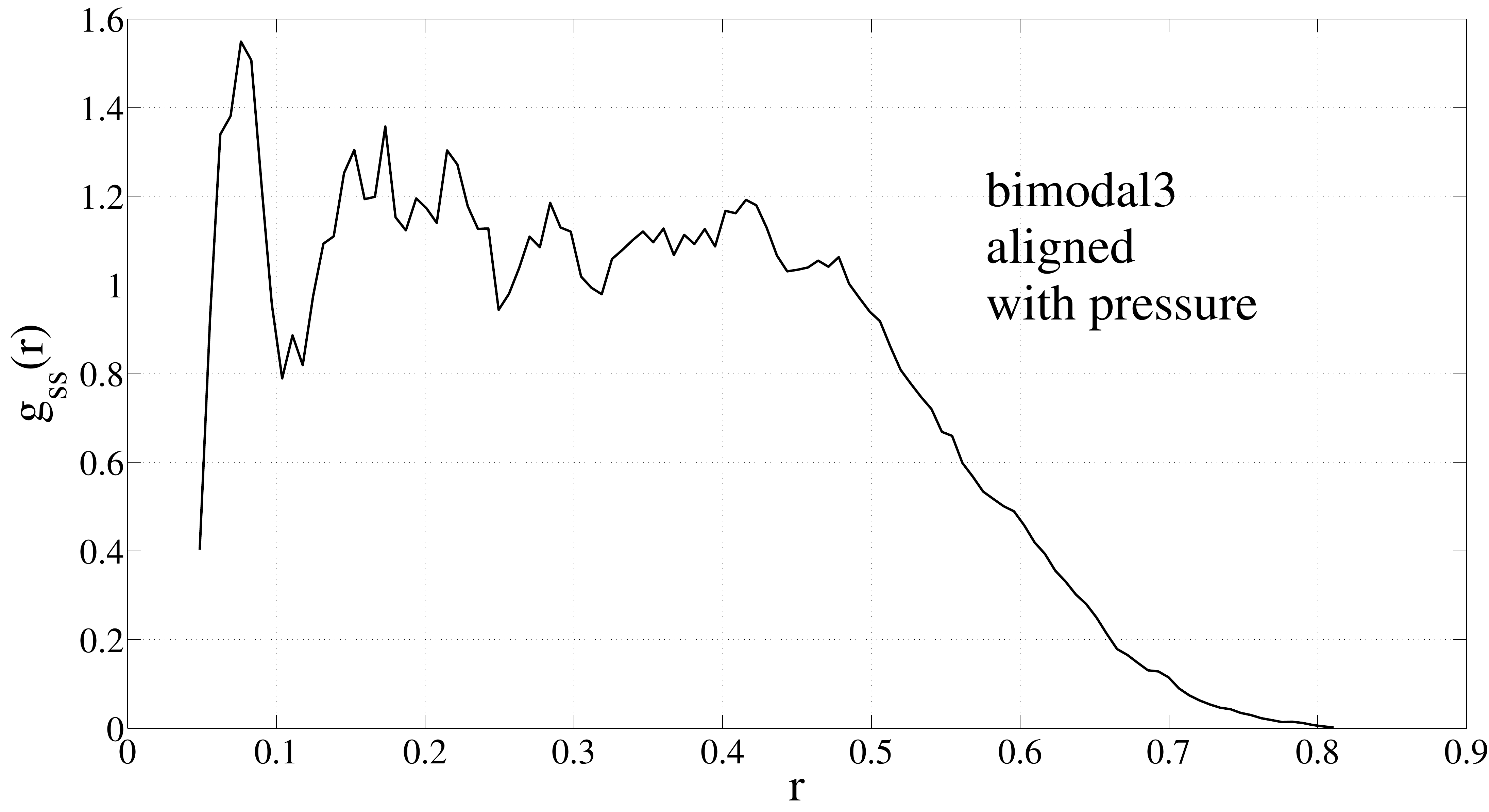}\\
\includegraphics[width=5.5cm]{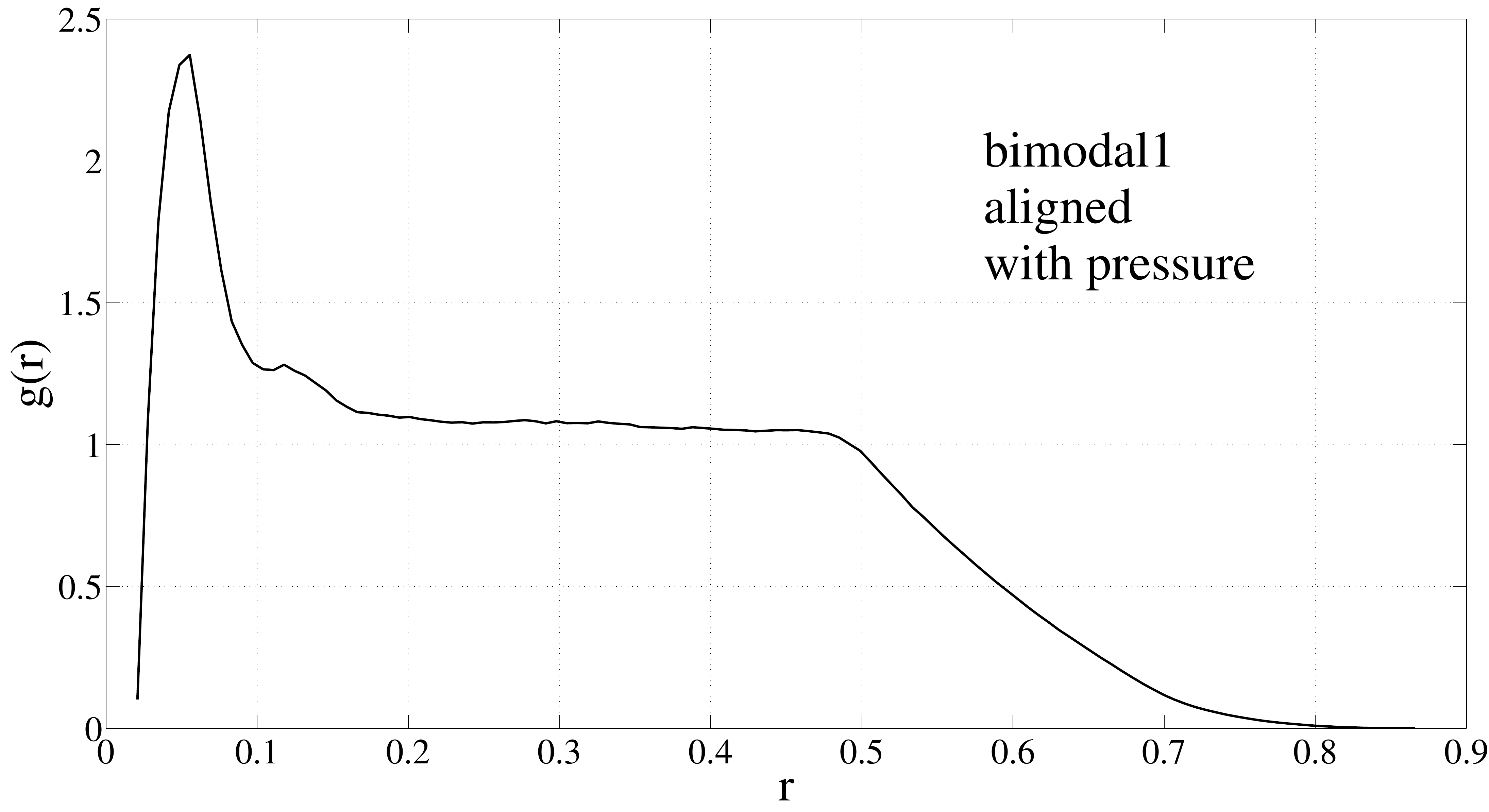}
\includegraphics[width=5.5cm]{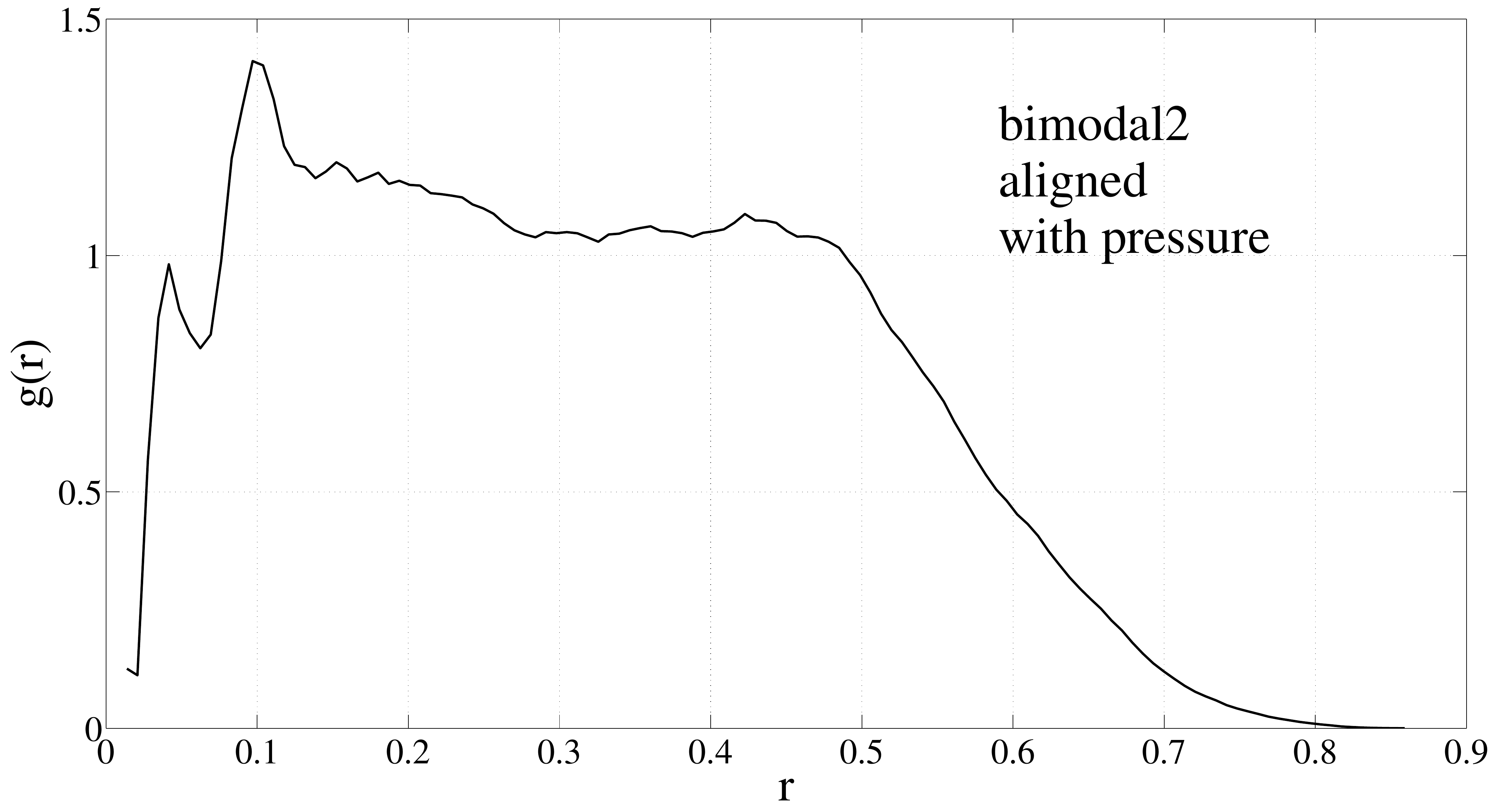}
\includegraphics[width=5.5cm]{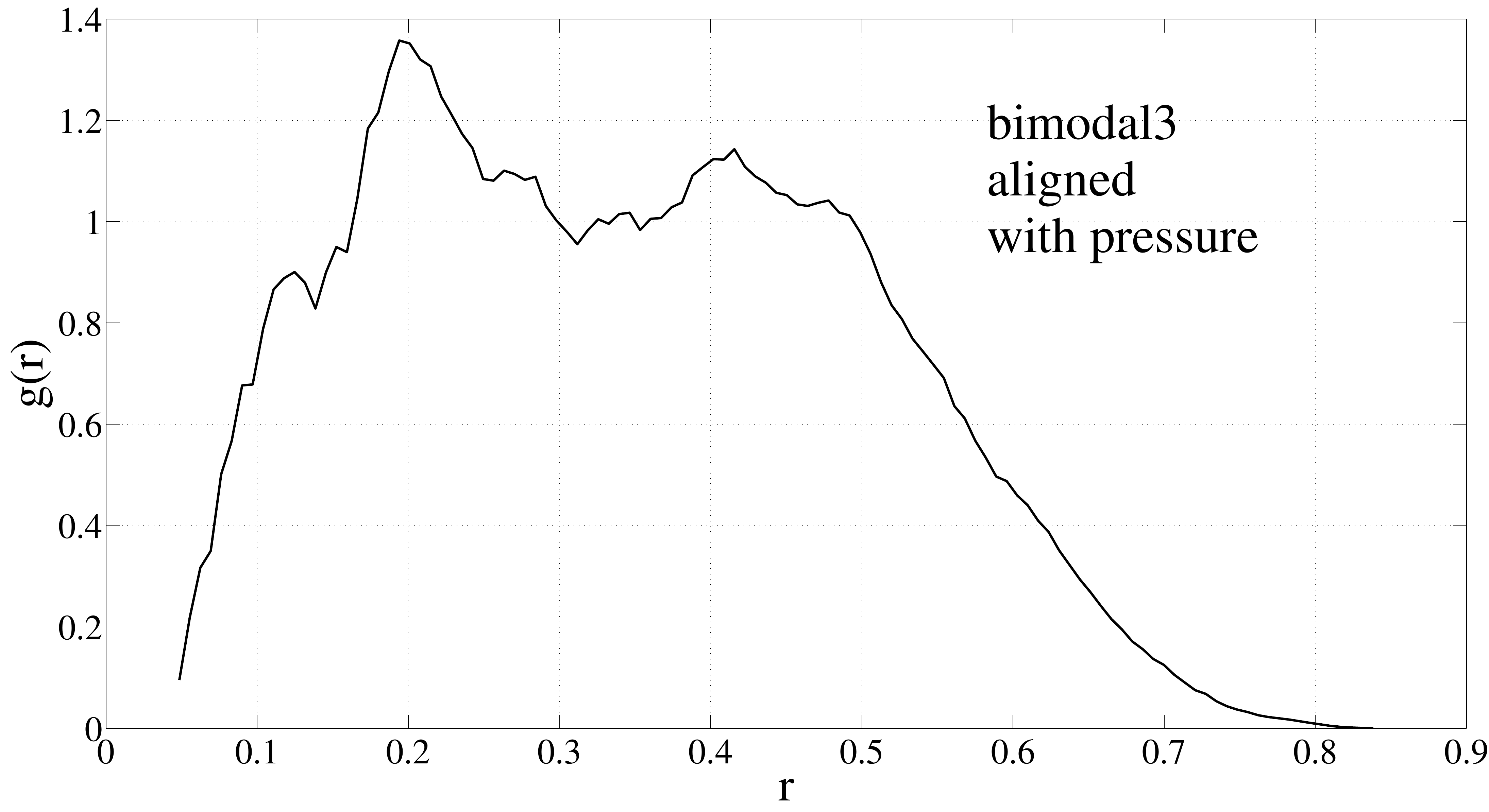}\\
\caption{Representative RDFs. For $\Delta r$ we choose $5/256$ (see Eq.~\ref{eq:rdf}).
\label{fig:rdf}}
\end{figure*}

\begin{table*}[th]
\begin{ruledtabular}
\begin{center}
\begin{tabular}{l|cccccc}
& $\mathrm{I}$ & $\mathrm{II}$ & $\mathrm{III}$ & $\mathrm{IV}$ & $\mathrm{V}$ & $\mathrm{VI}$\\
\hline
\multirow{3}*{bimodal1 (std)}
& $0$-$0$ & $0.5159$-$0.1329$ & $0$-$0$           & $0$-$0$           & $0.5209$-$0.1355$ & $0.5072$-$0.1287$ \\
& $0$-$0$ & $0$-$0$           & $0.0586$-$0.0017$ & $0.0588$-$0.0017$ & $0.0585$-$0.0017$ & $0.0590$-$0.0017$ \\
& $0$-$0$ & $0$-$0$           & $0$-$0$           & $0.0576$-$0.0017$ & $0$-$0$           & $0.0585$-$0.0017$ \\
\hline
\multirow{3}*{bimodal1 (press)} 
& $0$-$0$ & $0.5140$-$0.1321$ & $0$-$0$           & $0$-$0$           & $0.5142$-$0.1324$ & $0.5084$-$0.1294$ \\
& $0$-$0$ & $0$-$0$           & $0.0579$-$0.0017$ & $0.0577$-$0.0017$ & $0.0570$-$0.0016$ & $0.0585$-$0.0017$ \\
& $0$-$0$ & $0$-$0$           & $0$-$0$           & $0.0581$-$0.0017$ & $0$-$0$           & $0.0580$-$0.0017$ \\
\hline
\multirow{3}*{bimodal2 (std)} 
& $0$-$0$ & $0.4895$-$0.1206$ & $0$-$0$           & $0$-$0$           & $0.4761$-$0.1153$ & $0.4808$-$0.1182$\\
& $0$-$0$ & $0$-$0$           & $0.0571$-$0.0016$ & $0.0565$-$0.0016$ & $0.0550$-$0.0015$ & $0.0564$-$0.0016$\\
& $0$-$0$ & $0$-$0$           & $0$-$0$           & $0.0553$-$0.0015$ & $0$-$0$           & $0.0566$-$0.0016$\\
\hline
\multirow{3}*{bimodal2 (press)} 
& $0$-$0$ & $0.4780$-$0.1146$ & $0$-$0$           & $0$-$0$           & $0.4884$-$0.1198$ & $0.4935$-$0.1218$ \\
& $0$-$0$ & $0$-$0$           & $0.0584$-$0.0017$ & $0.0586$-$0.0017$ & $0.0553$-$0.0015$ & $0.0553$-$0.0015$ \\
& $0$-$0$ & $0$-$0$           & $0$-$0$           & $0.0585$-$0.0017$ & $0$-$0$           & $0.0565$-$0.0016$ \\
\hline
\multirow{3}*{bimodal3 (std)}
& $0$-$0$ & $0.5001$-$0.1248$ & $0$-$0$           & $0$-$0$           & $0.4896$-$0.1227$ & $0.4707$-$0.1114$\\
& $0$-$0$ & $0$-$0$           & $0.0579$-$0.0017$ & $0.0533$-$0.0014$ & $0.0493$-$0.0012$ & $0.0568$-$0.0016$\\
& $0$-$0$ & $0$-$0$           & $0$-$0$           & $0.0570$-$0.0016$ & $0$-$0$           & $0.0583$-$0.0017$\\
\hline
\multirow{3}*{bimodal3 (press)}  
& $0$-$0$ & $0.4870$-$0.1186$ & $0$-$0$           & $0$-$0$           & $0.4694$-$0.1127$ & $0.4577$-$0.1077$\\
& $0$-$0$ & $0$-$0$           & $0.0545$-$0.0015$ & $0.0566$-$0.0016$ & $0.0538$-$0.0016$ & $0.0590$-$0.0017$\\
& $0$-$0$ & $0$-$0$           & $0$-$0$           & $0.0524$-$0.0014$ & $0$-$0$ & $0.0556$-$0.0015$          \\
\hline
\multirow{3}*{granular (std)} 
& $0$-$0$ & $0.4949$-$0.1200$ & $0$-$0$            & $0$-$0$           & $0.4870$-$0.1222$ & $0.4779$-$0.1185$ \\
& $0$-$0$ & $0$-$0$           & $0.0555$-$0.0015$  & $0.0571$-$0.0016$ & $0.0531$-$0.0014$ & $0.0532$-$0.0014$ \\
& $0$-$0$ & $0$-$0$           & $0$-$0$            & $0.0548$-$0.0015$ & $0$-$0$           & $0.0579$-$0.0016$ \\
\hline
\multirow{3}*{granular (press)} 
& $0$-$0$ & $0.5175$-$0.1336$ & $0$-$0$           & $0$-$0$           & $0.4617$-$0.1126$ & $0.5162$-$0.1312$ \\
& $0$-$0$ & $0$-$0$           & $0.0579$-$0.0016$ & $0.0537$-$0.0014$ & $0.0609$-$0.0018$ & $0.0601$-$0.0018$ \\
& $0$-$0$ & $0$-$0$           & $0$-$0$           & $0.0577$-$0.0016$ & $0$-$0$           & $0.0570$-$0.0016$ \\
\hline
\multirow{3}*{fine (std)} 
& $0$-$0$ & $0.5227$-$0.1366$ & $0$-$0$           & $0$-$0$           & $0.5162$-$0.1332$ & $0.5146$-$0.1325$ \\
& $0$-$0$ & $0$-$0$           & $0.0589$-$0.0017$ & $0.0581$-$0.0017$ & $0.0581$-$0.0017$ & $0.0581$-$0.0017$ \\
& $0$-$0$ & $0$-$0$           & $0$-$0$           & $0.0582$-$0.0017$ & $0$-$0$           & $0.0576$-$0.0017$ \\
\hline
\multirow{3}*{fine (press)} 
& $0$-$0$ & $0.5200$-$0.1351$ & $0$-$0$           & $0$-$0$           & $0.5162$-$0.1331$ & $0.5095$-$0.1299$ \\
& $0$-$0$ & $0$-$0$           & $0.0589$-$0.0017$ & $0.0581$-$0.0017$ & $0.0578$-$0.0017$ & $0.0587$-$0.0017$ \\
& $0$-$0$ & $0$-$0$           & $0$-$0$           & $0.0574$-$0.0016$ & $0$-$0$           & $0.0587$-$0.0017$ \\
\end{tabular}
\end{center}
\caption{\label{tab:angles}
Average value and variance of the variables $\Delta\theta$ for the case
studies of Tabs.~\ref{tab:powders} and \ref{tab:case}. For every mixture, the first row
refers to $\Delta\theta_{z}$ followed by its variance, the second to $\Delta\theta_{x}$ and its variance,
and the third to $\Delta\theta_{y}$ and its variance. The abbreviations std and press mean with no pressure
and with pressure, respectively.}
\end{ruledtabular}
\end{table*}

In Tab.~\ref{tab:filling}, the packing fractions $\phi$ reached for the different powders and orientational
characteristics are reported. In general, the application of pressure leads to denser packings. We
ascribe the unique exception to statistical fluctuations. Indeed,
for the granular powder of type \textrm{IV} with pressure, average over five realizations gives $\phi=0.422$,
in this case above the value for the same system with no applied pressure. For comparative purposes 
relative to the different orientational properties, Fig.~\ref{fig:layers} shows the sections of four final 
configurations for the powder bimodal1 with an applied pressure. In this case, the highest packing values 
are reached since the smaller particles can better fill the irregular interstices between larger blocks. 
Figure \ref{fig:3d} shows the $3$D representation of the packing structure when the role of pressure 
is more effective (bimodal2 powder of type \textrm{I}). In general, orientational disorder lowers the packing
density in particular if misalignment is allowed (see Fig.~\ref{fig:v2phi}). Fine powders lead to looser random packings. 
It turns out that the highest volume fractions can be achieved with more polydisperse size distributions 
by starting from the addition of larger blocks in the absence of orientational disorder. The role of smaller 
particles is to fill progressively narrower interstices. In this way, it is possible to simulate the
infiltration of finer powders between stuck particles and eventually their aggregation. Table \ref{tab:density}
further details the results for the packing structures of all powders.

It is interesting to look at how the packing fraction progresses in the course of time. 
Figure \ref{fig:t2phi} shows the cases of granular and fine powders of type \textrm{I} (aligned case).
With pressure, the volume fraction follows a power law $\phi\sim t^{\gamma}$ over one more decade.
Furthermore, at the end, these simulations are faster. For other orientational characteristics, it is
found that $\phi$ displays an analogous behavior. In the case of bimodal powders, we also find similar
curves for the addition of smaller blocks.

In order to gain insight into the structural properties of the final configurations of the generated
powders, we calculate the radial distribution functions (RDFs) between types of blocks. The usual
definition of the RDF $g(r)$ is \cite{rdf_book}
\begin{equation}
\frac{N_{1}}{V}g(r)4\pi r^{2}\Delta r=S(r)\ .
\label{eq:rdf}
\end{equation}
$V=1$ is the volume of the domain. $S(r)$ is the average number of particles of type $2$ falling
within a spherical shell of radius $r$ and width $\Delta r$ centered around the particles of type 1
present in number $N_{1}$. (Precisely, the distance between two blocks is determined using their centers of mass.) 
As a consequence, the RDF is a measure of the average radial dependence of the density of particles of type 2 
around those of type 1. In the following, given a powder, with the notation $g_{\mathrm{l}\mathrm{s}}(r)$ we intend 
the RDF of smaller particles around larger ones. We employ similar notations for the other cases and, with no upper-script, 
the RDF is among all particles. Figure \ref{fig:rdf} shows different RDFs in several cases. 
In the first row, the main peak of $g_{\mathrm{l}\mathrm{l}}$ is determined by adjacent in-plane blocks.
The first peak around $r= 0.1$ is due to the elongated shape of the blocks. It is lower than the second
peak because a block can have indicatively twice more in-plane neighbors. In the absence of pressure
(second row), it is seen that the peaks are lower and slightly broader, indicating that these structures
are less compact. Of course, these remarks hold also for the other bimodal powders. The third row of Fig.~\ref{fig:rdf}
shows $g_{\mathrm{l}\mathrm{s}}$ for different mixtures. Mainly, as the volume ratio gets smaller, there appears more
distinctly a peak at $r\approx 0.1$. This means that the contribution of out-of-plane small particles
become statistically more significant. 
For the powder bimodal3, the last peak around $r=0.4$ is related
to the first peak of $g_{\mathrm{l}\mathrm{l}}$ and corresponds to pairs of large-small blocks separated by a large one.
For the bimodal1 powder, the first peak of $g_{\mathrm{s}\mathrm{s}}$ always
occurs before $r=0.1$ (fourth row). It follows that this peak is due to the small particles filling
the voids between larger blocks. The same RDF for the fine powder is similar (fifth row).  
The absence of interstices between
larger blocks has no consequence because the small particles tend to fill densely the voids. Indeed,
for the other bimodal powders there appear additional peaks after $r=0.1$, more distinctly for the lowest
volume ratio (bimodal3 powder). The RDF among all particles for the bimodal1 powder (last row) tells us 
that the statistics is dominated by the small particles relatively close to each other determining the 
first peak. In the other cases, the arrangement of smaller blocks is progressively similar to that of 
larger ones and there appear two peaks for $r\geq 0.1$ for the reasons explained above (see first row).

Interestingly, Fig.~\ref{fig:layers} suggests that the orientation of larger blocks might be correlated.
In order to better understand this point, we carry out a statistics of the rotation angles.
Let $P$ be the number of possible pairs of blocks. We introduce the average 
quantities
\begin{eqnarray*}
&&\langle\Delta\theta_{z} \rangle=\frac{1}{P}\sum_{i>j}^{P}\mid\theta_{z}(i)-\theta_{z}(j)\mid\ ;\\
&&\langle(\Delta\theta_{z})^{2} \rangle=\frac{1}{P}\sum_{i>j}^{P}[\theta_{z}(i)-\theta_{z}(j)]^{2}\ . 
\end{eqnarray*}
$\theta_{z}(i)$ is the rotation angle of the $i$-th block. Similar average values are defined for
the tilt angles $\theta_{x}$ and $\theta_{y}$. Besides the average values $\langle\Delta\theta\rangle$,
of interest are also the variances $\langle(\Delta\theta)^{2}\rangle-\langle\Delta\theta\rangle^{2}$.
In Tab.~\ref{tab:angles} these quantities are given in all cases addressed by simulations.
The blocks are expected to have the same orientation to a higher degree when the quantities
$\langle\Delta\theta\rangle$ are small. It is important to remark that $\langle\Delta\theta_{z}\rangle$
can be at most $\pi/2$. It is found that this quantity varies in the interval $0.4617-0.5227$, 
i.e.~$26^{\circ}-30^{\circ}$. As a result, from this overall statistics it clearly arises that orientational
ordering is present in the internal packing structures. However, it seems difficult to discriminate between 
and compare the different cases on the basis of this method.

\begin{figure*}[ht]
\includegraphics[width=5.5cm]{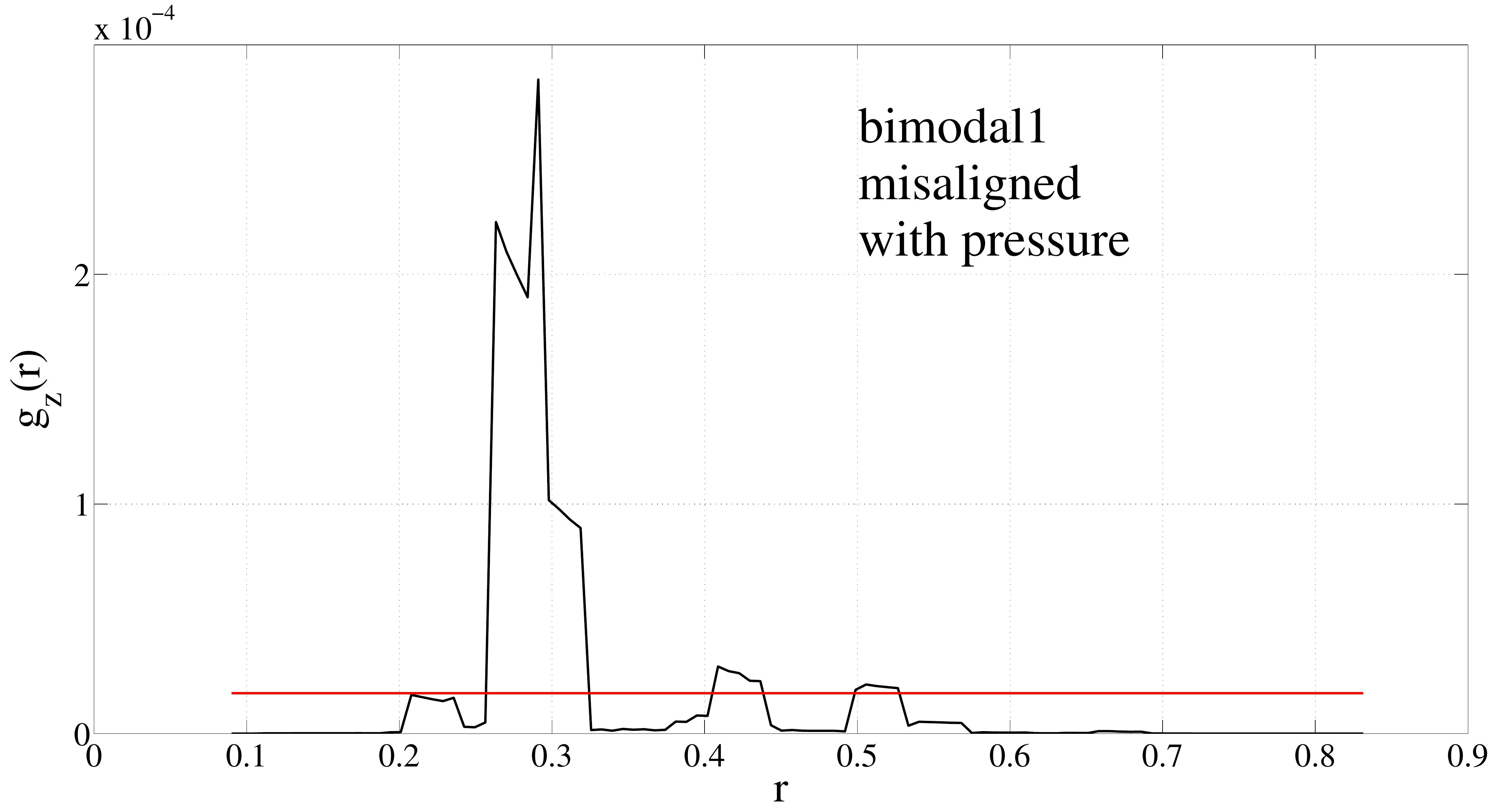}
\includegraphics[width=5.5cm]{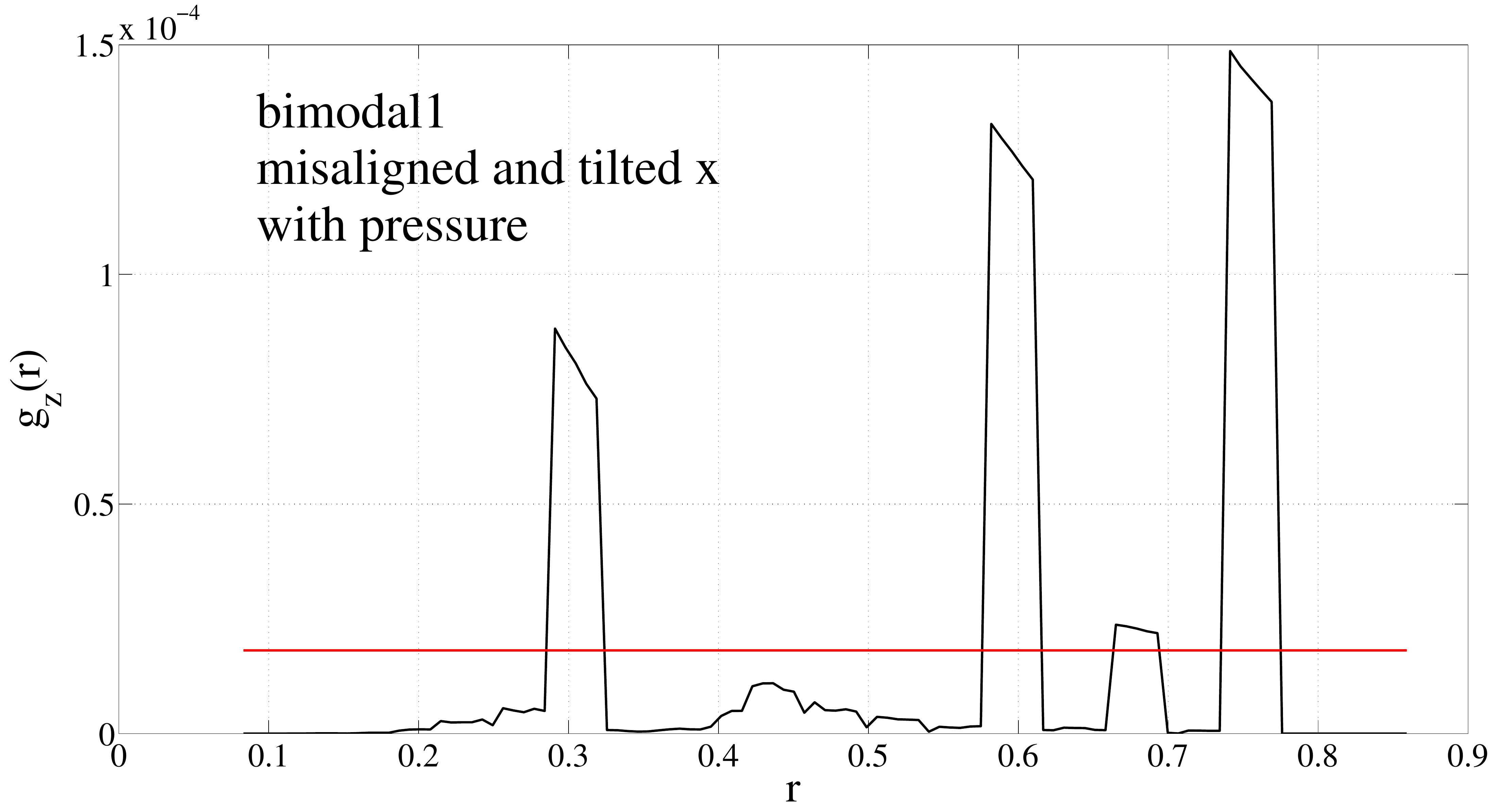}
\includegraphics[width=5.5cm]{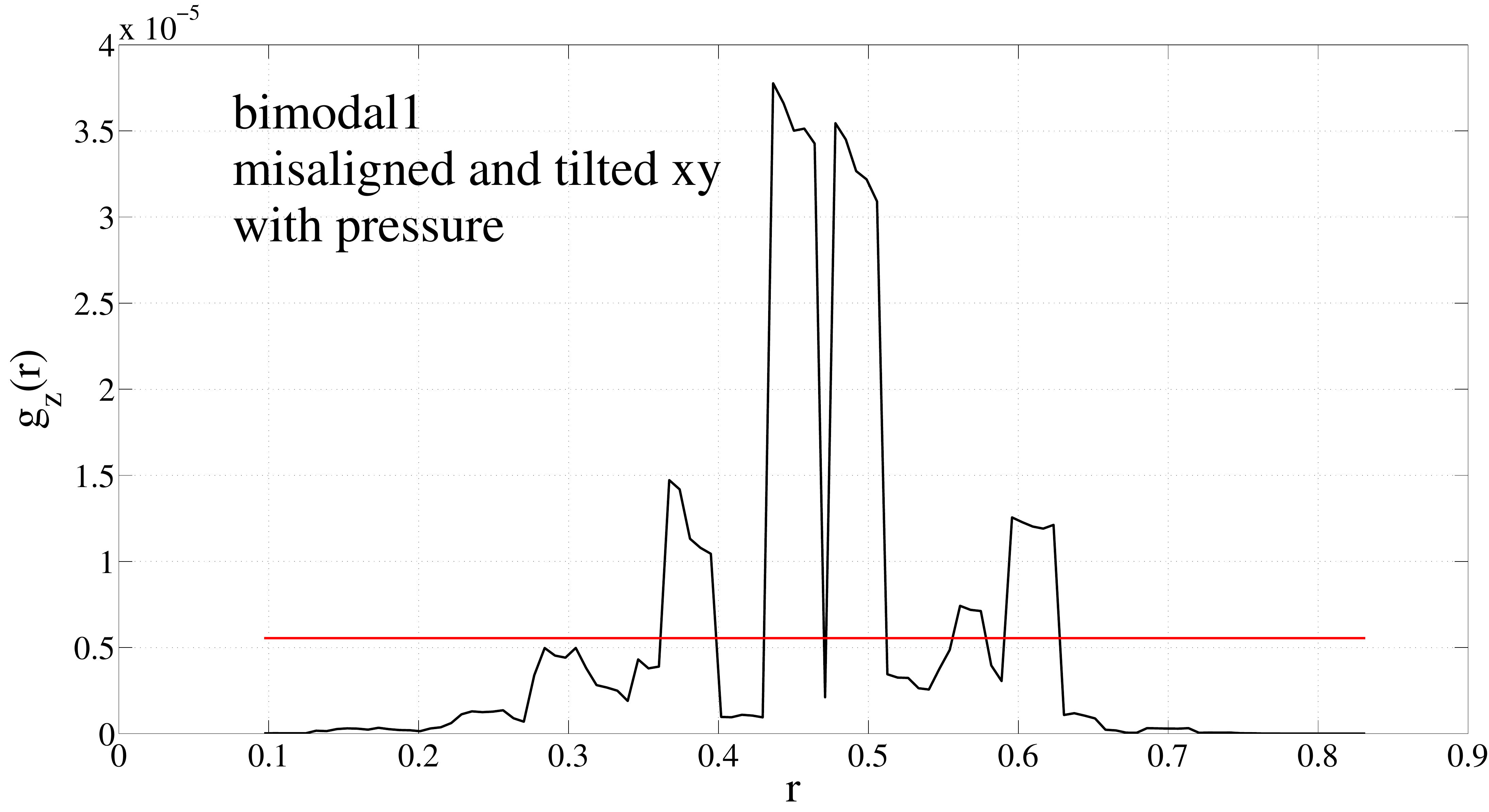}\\
\includegraphics[width=5.5cm]{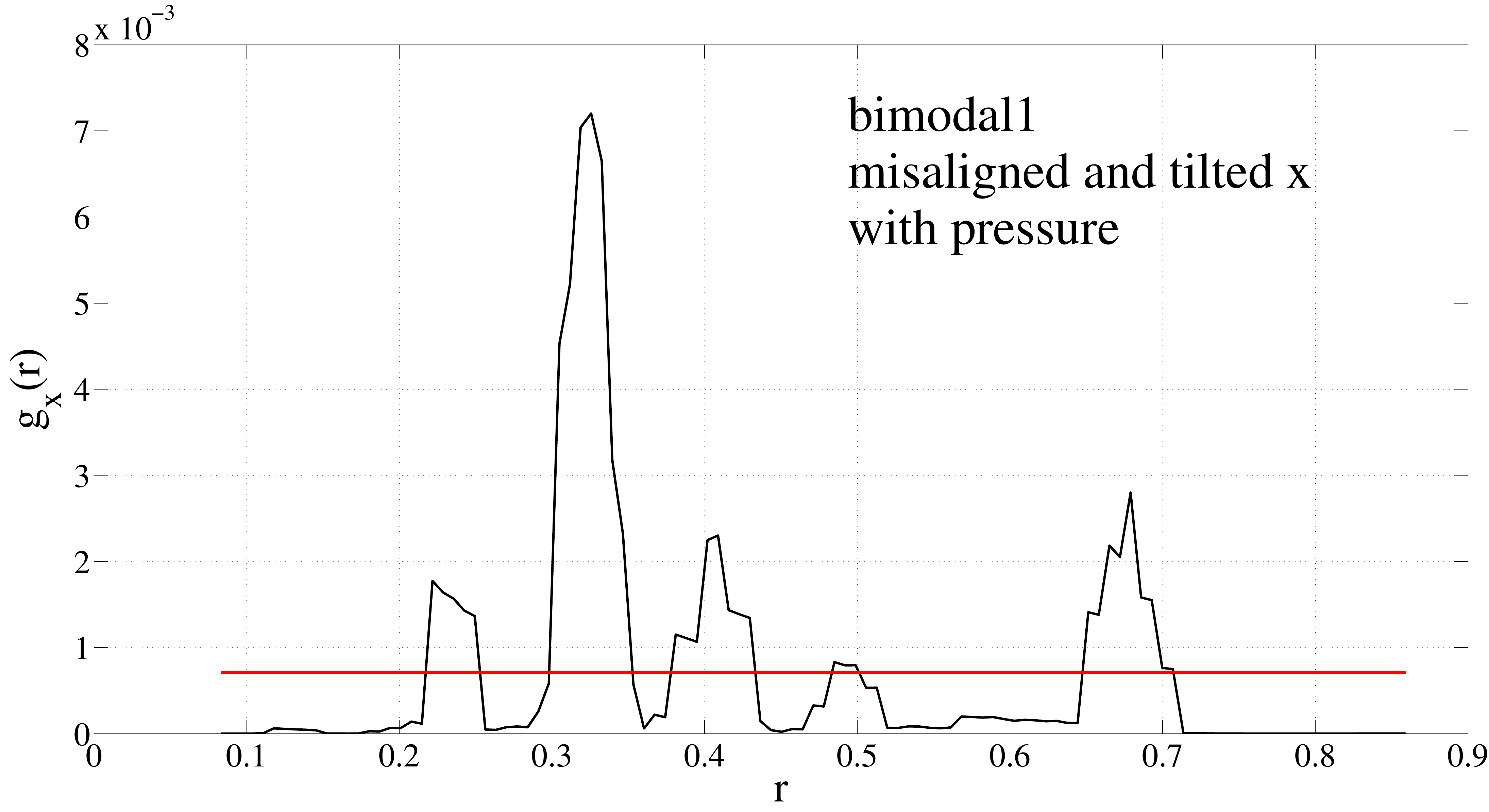}
\includegraphics[width=5.5cm]{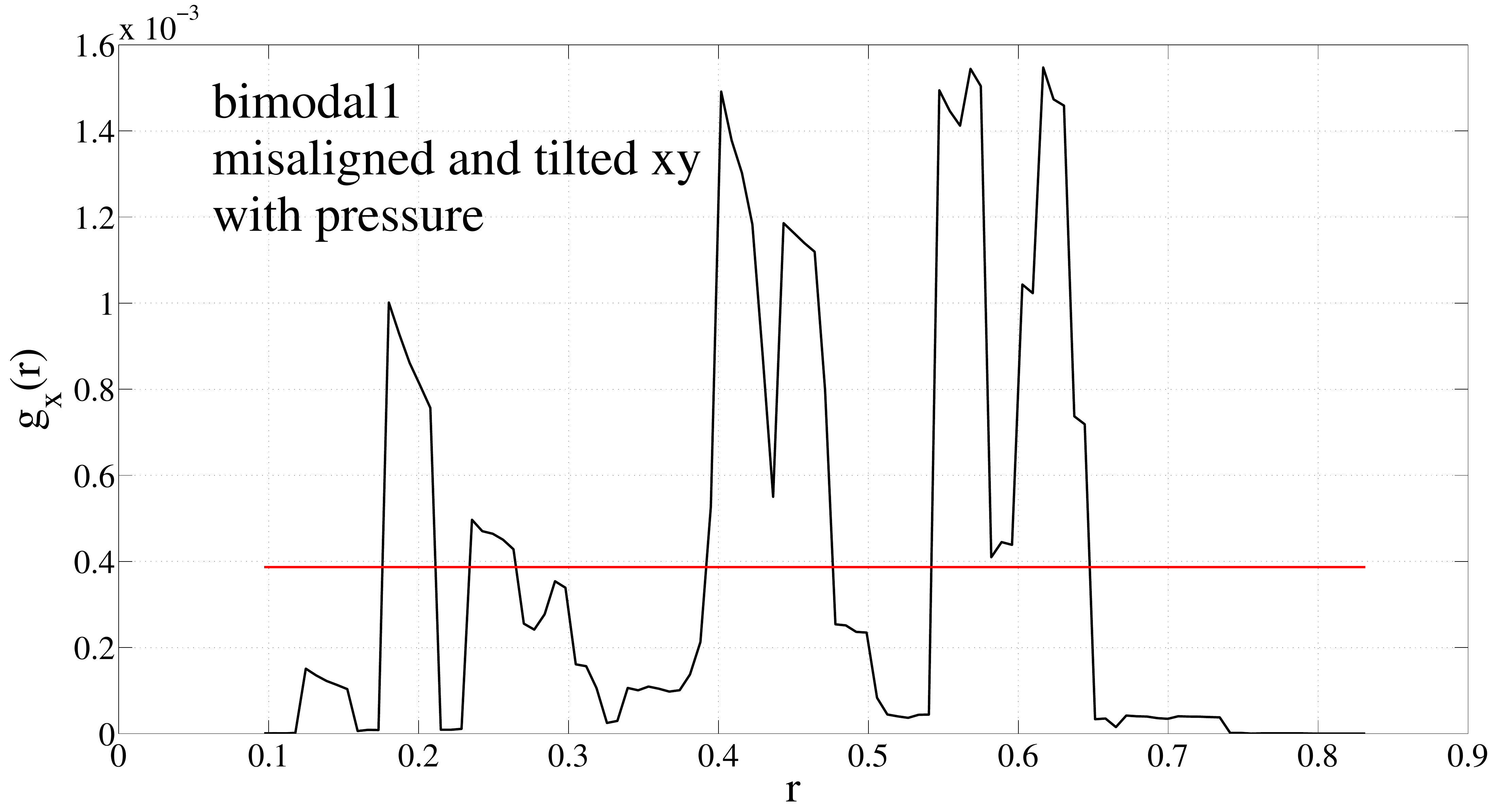}
\includegraphics[width=5.5cm]{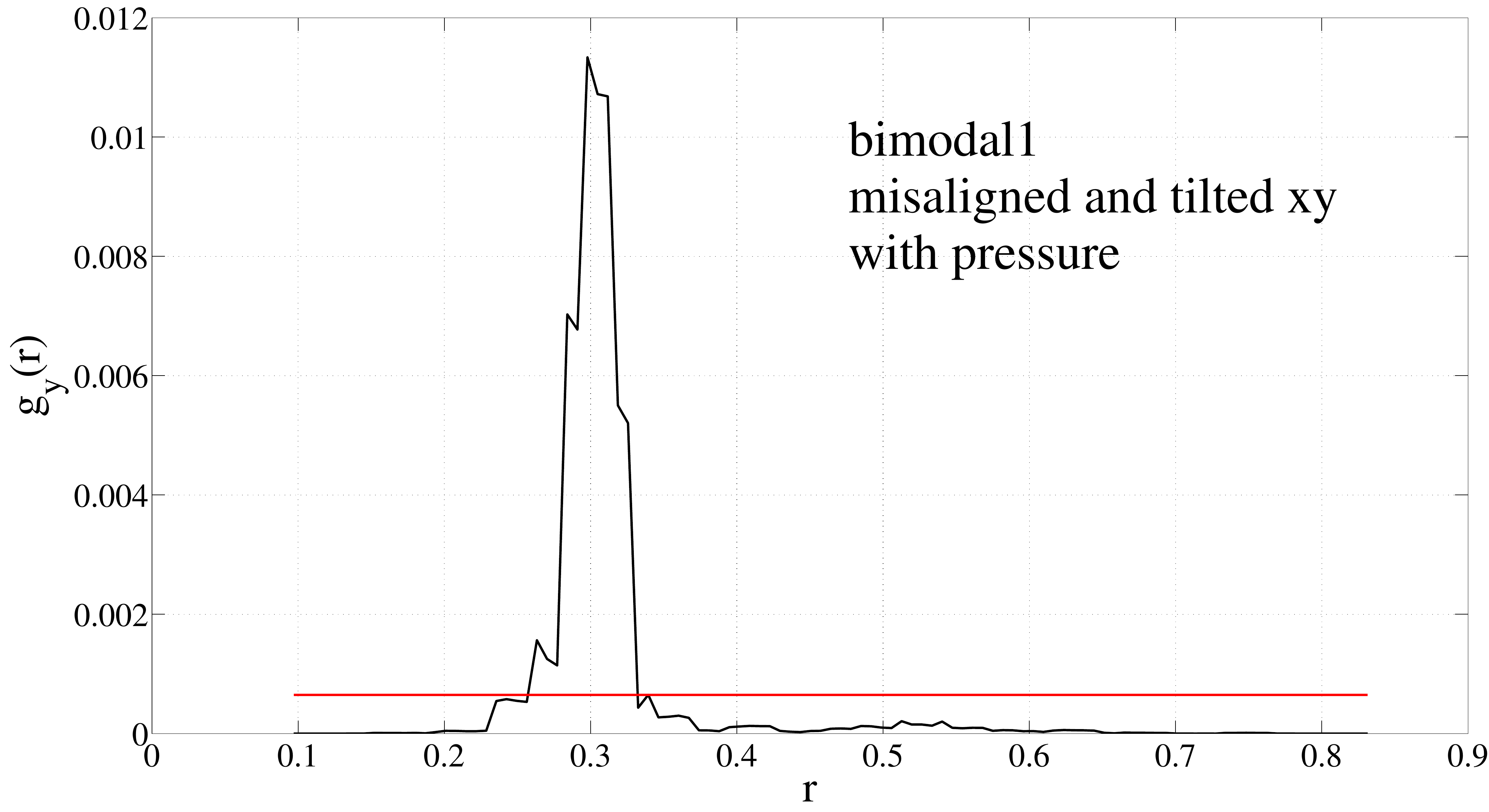}\\
\includegraphics[width=5.5cm]{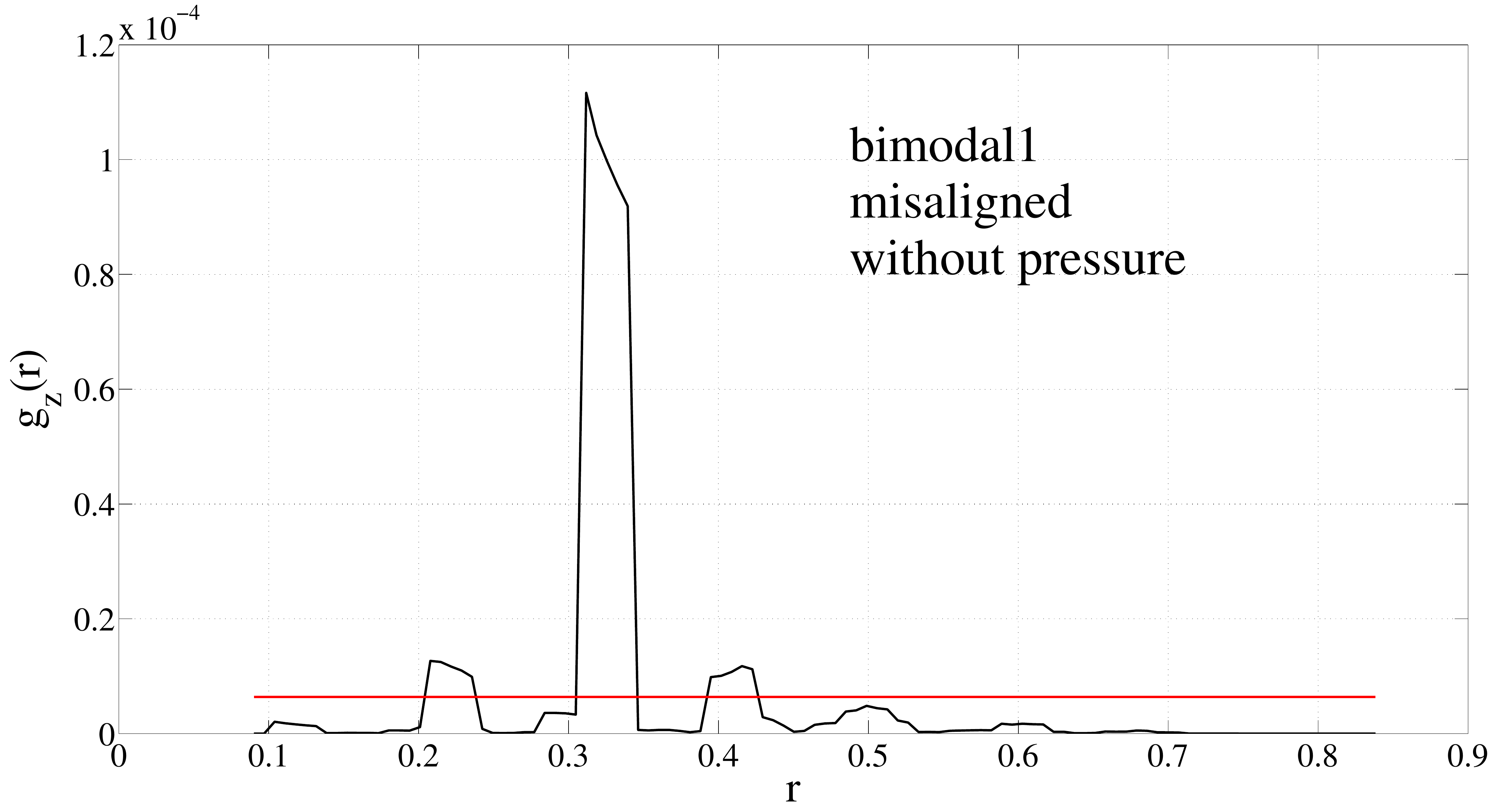}
\includegraphics[width=5.5cm]{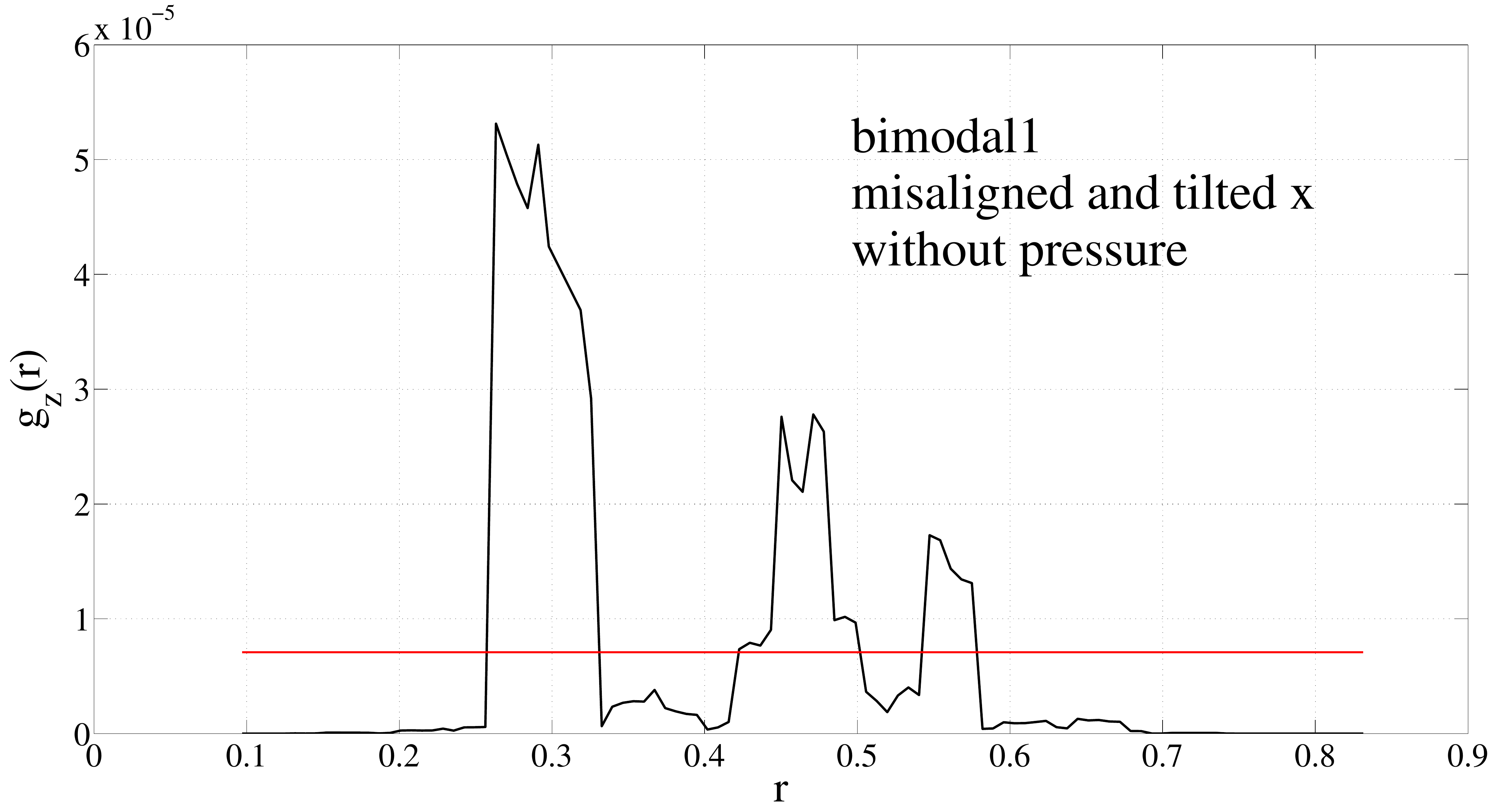}
\includegraphics[width=5.5cm]{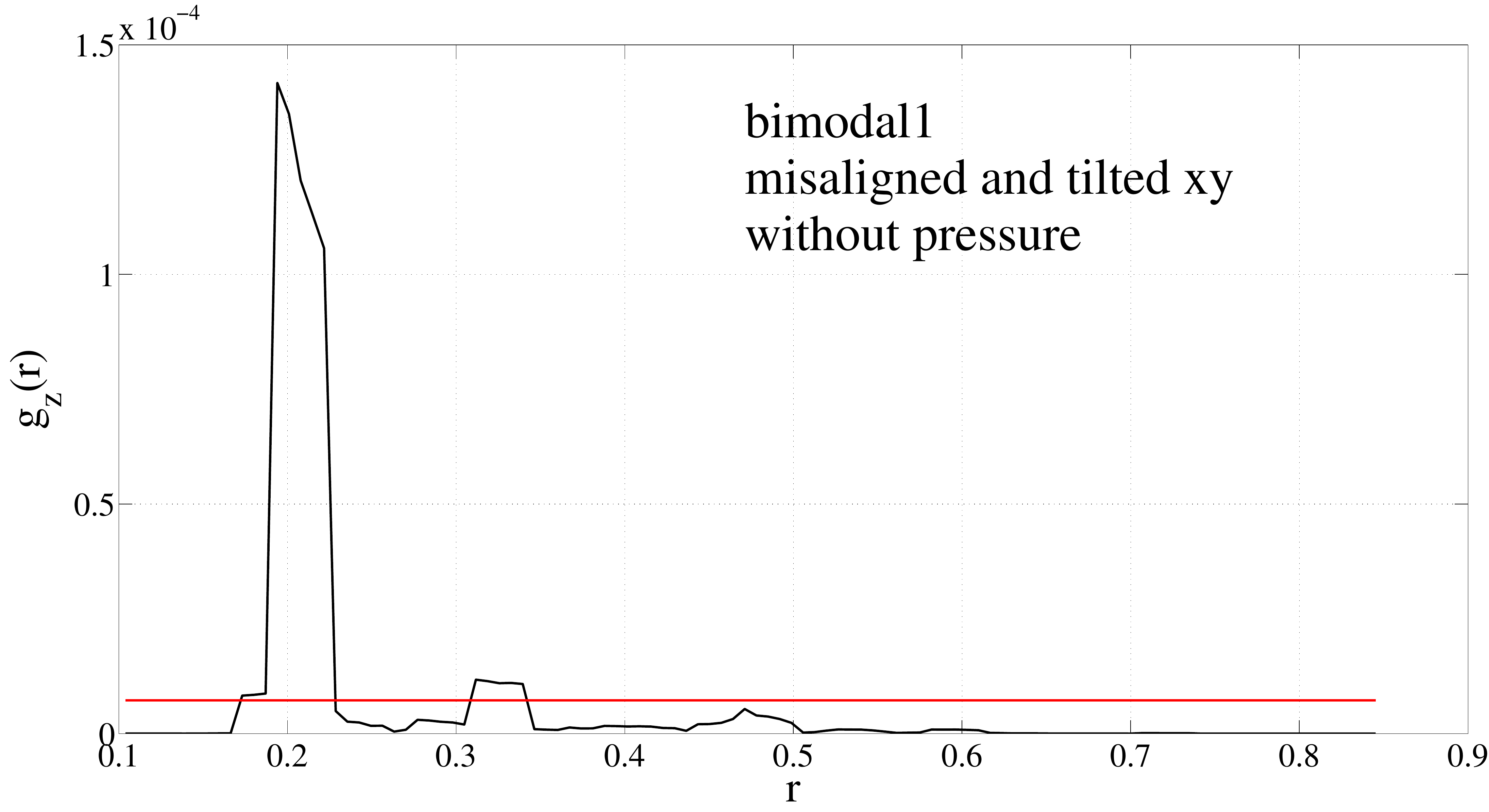}\\
\includegraphics[width=5.5cm]{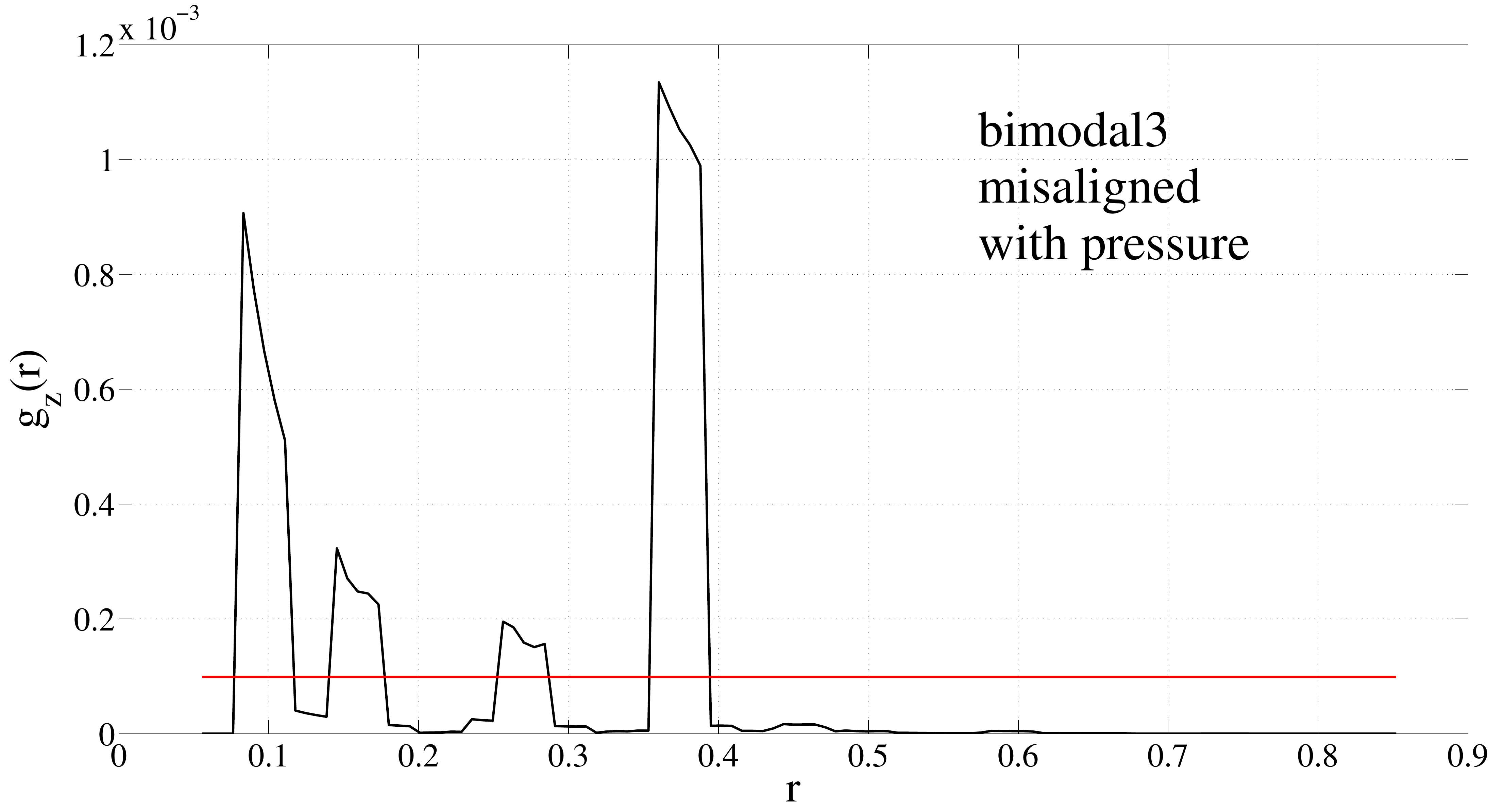}
\includegraphics[width=5.5cm]{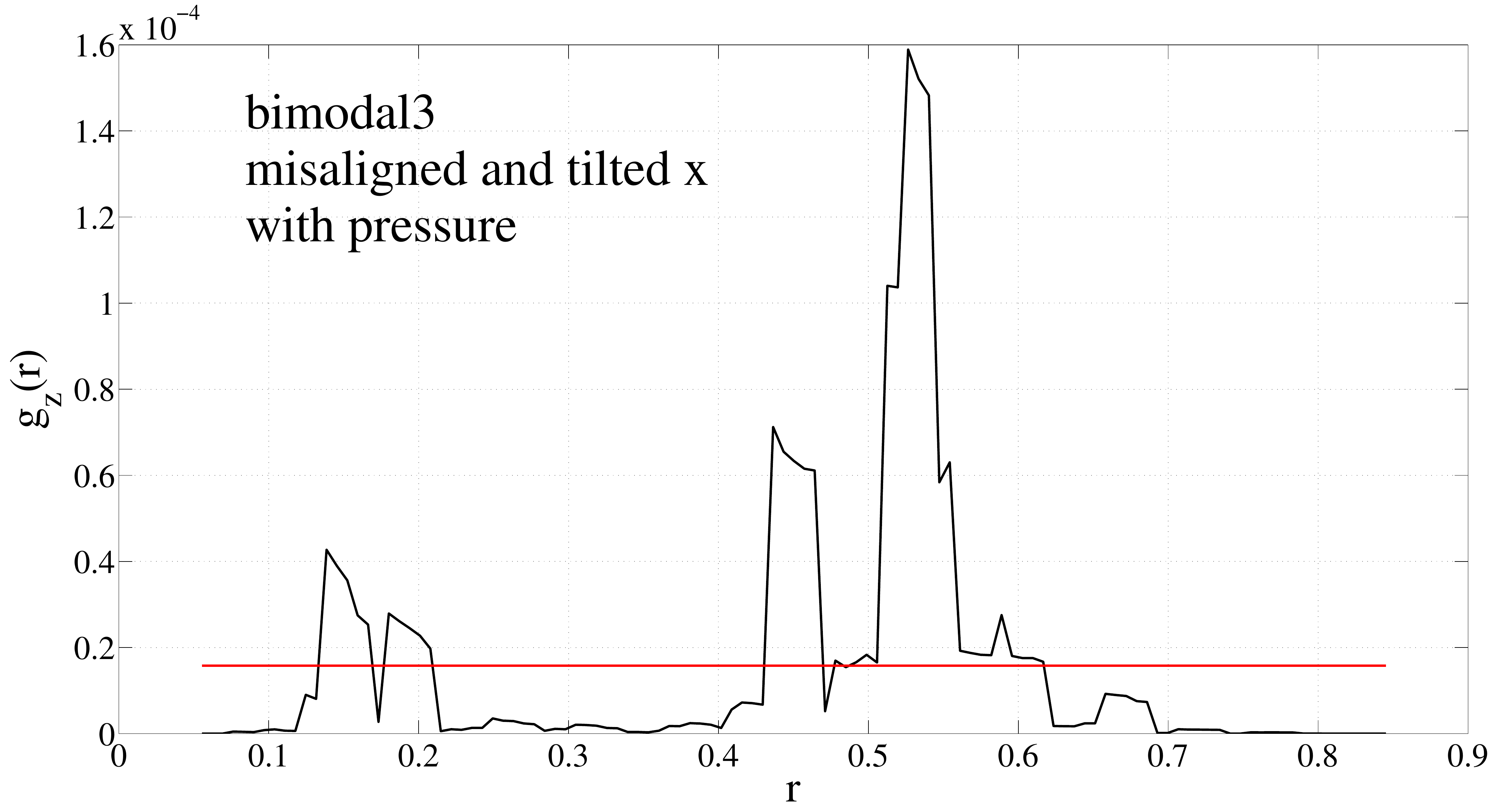}
\includegraphics[width=5.5cm]{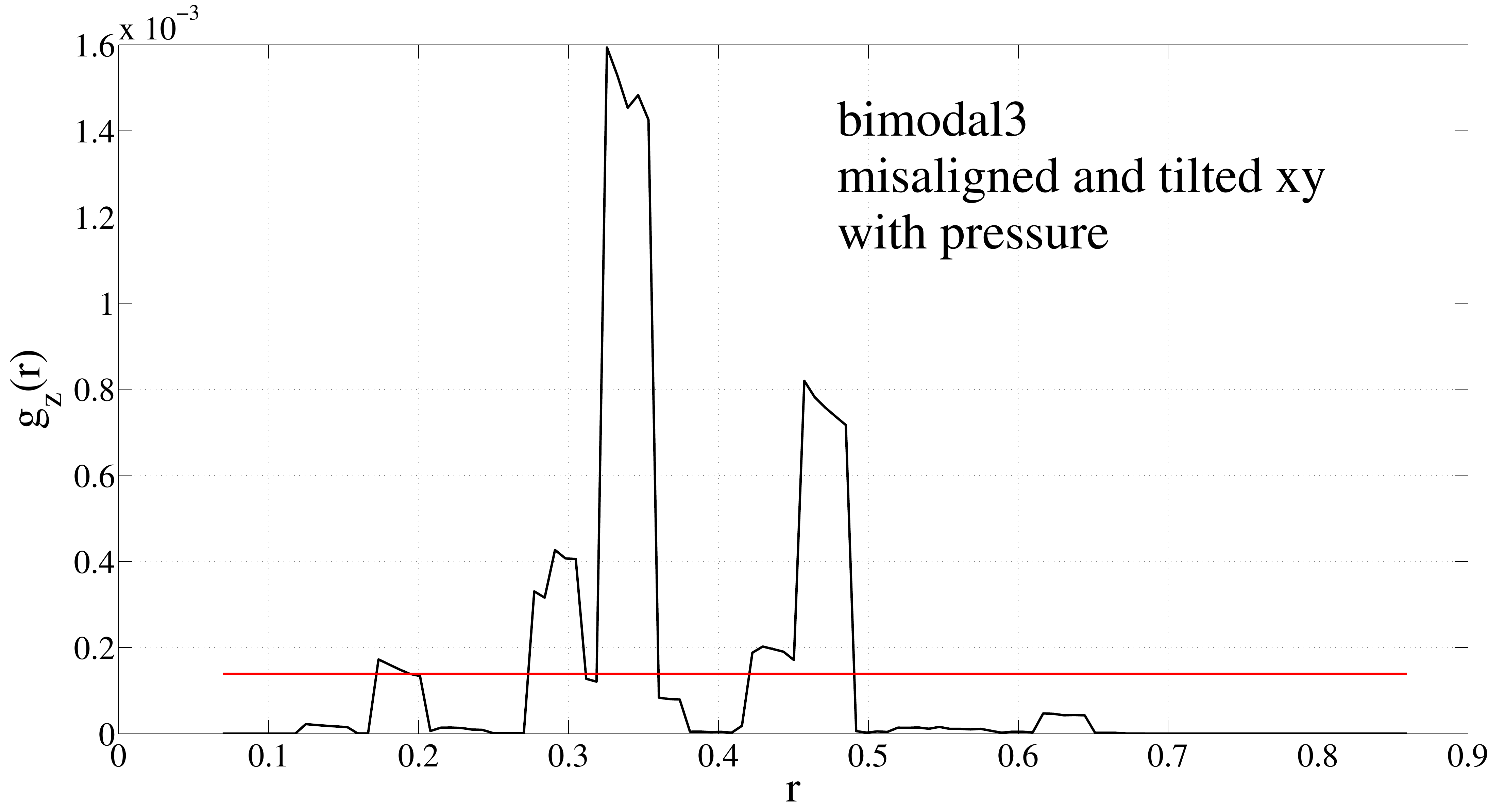}\\
\includegraphics[width=5.5cm]{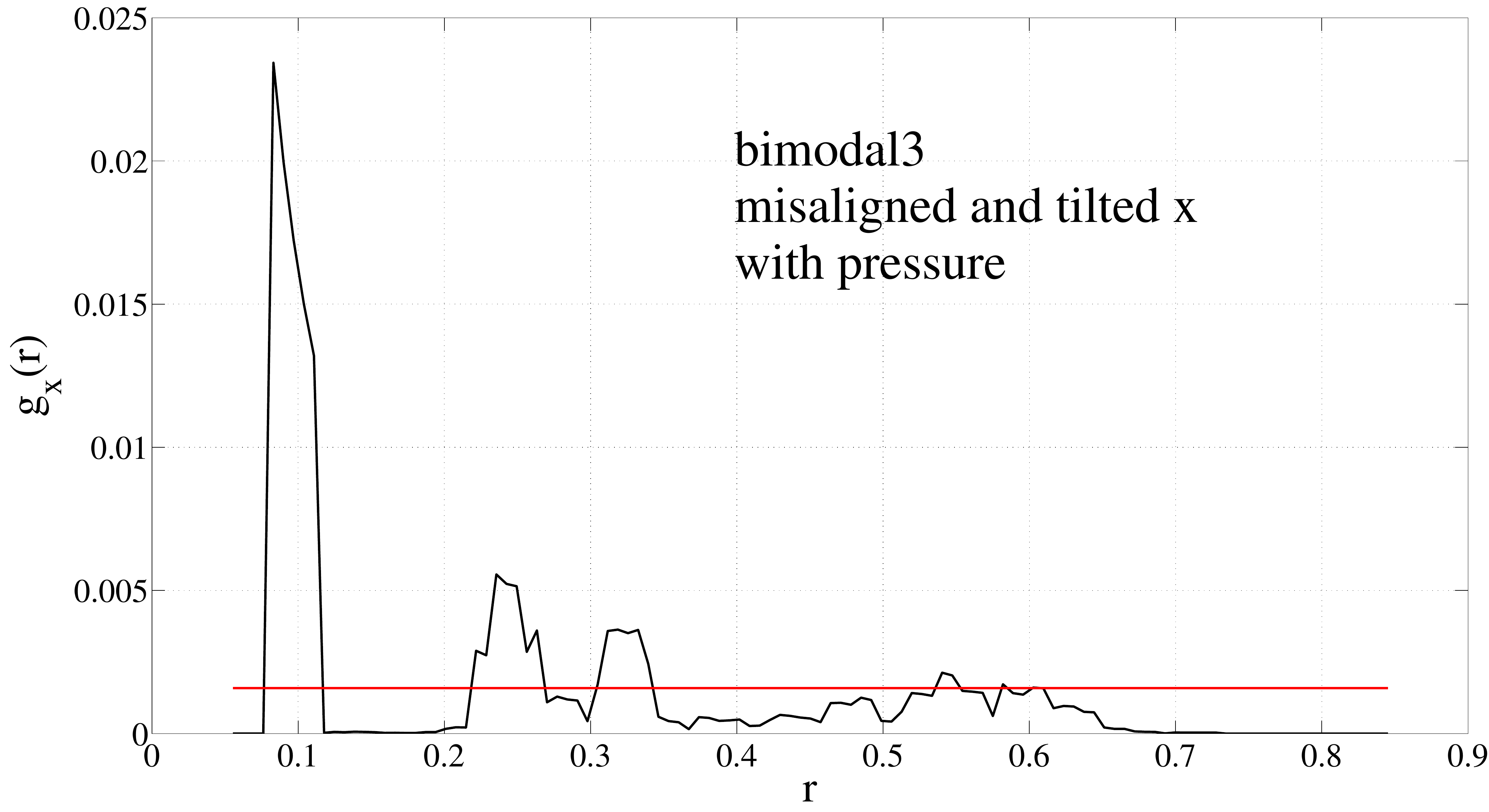}
\includegraphics[width=5.5cm]{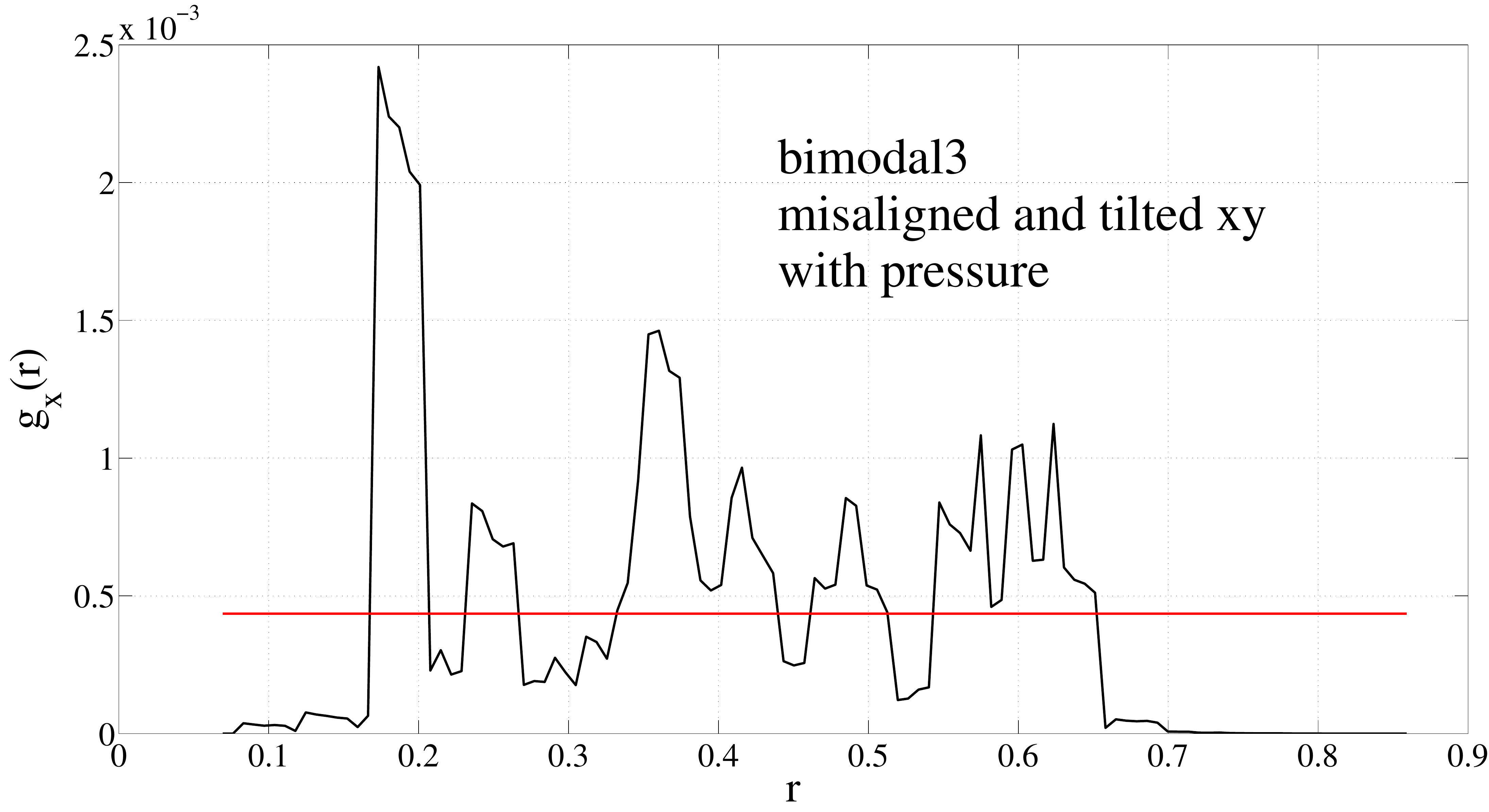}
\includegraphics[width=5.5cm]{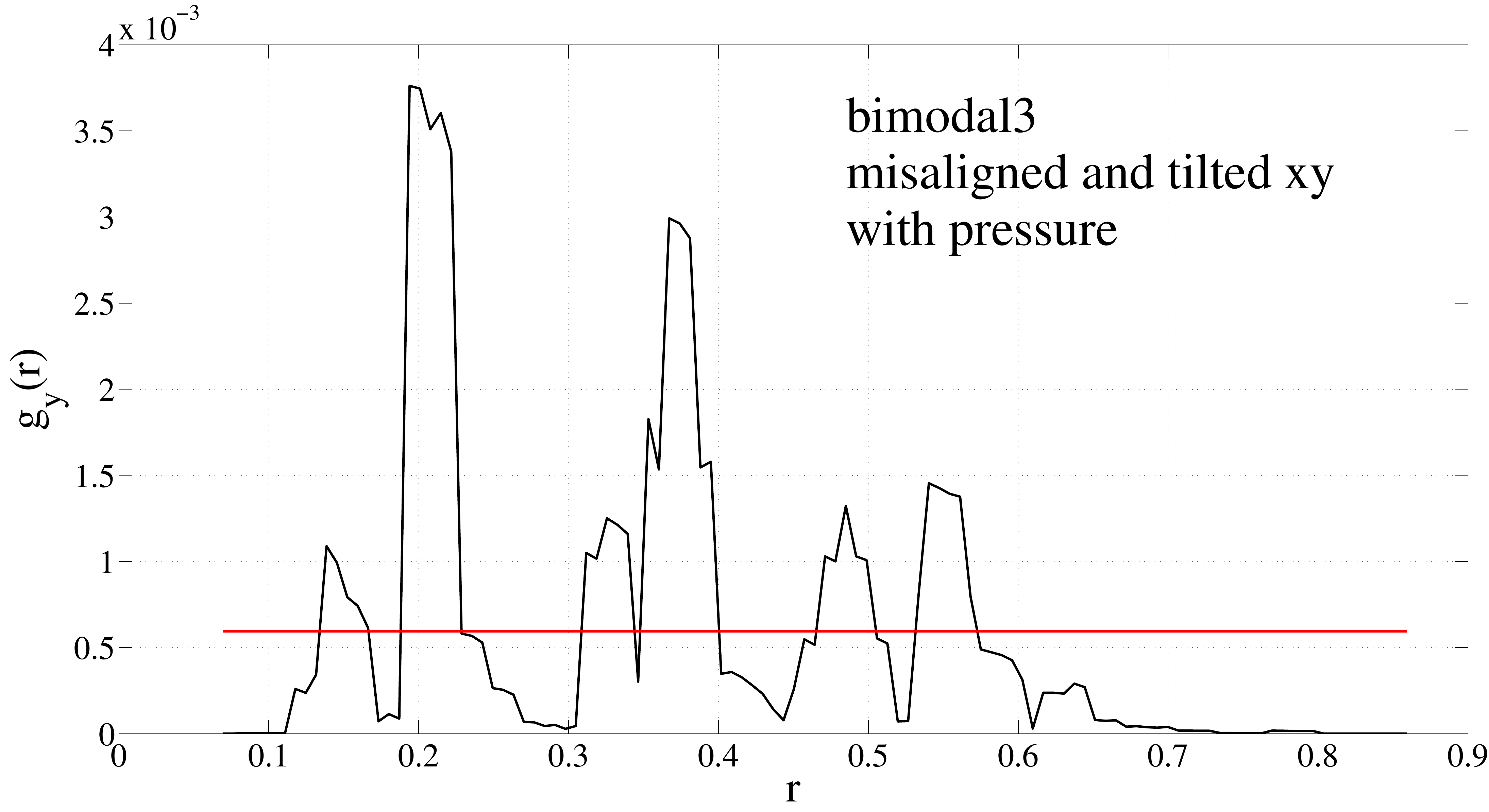}\\
\includegraphics[width=5.5cm]{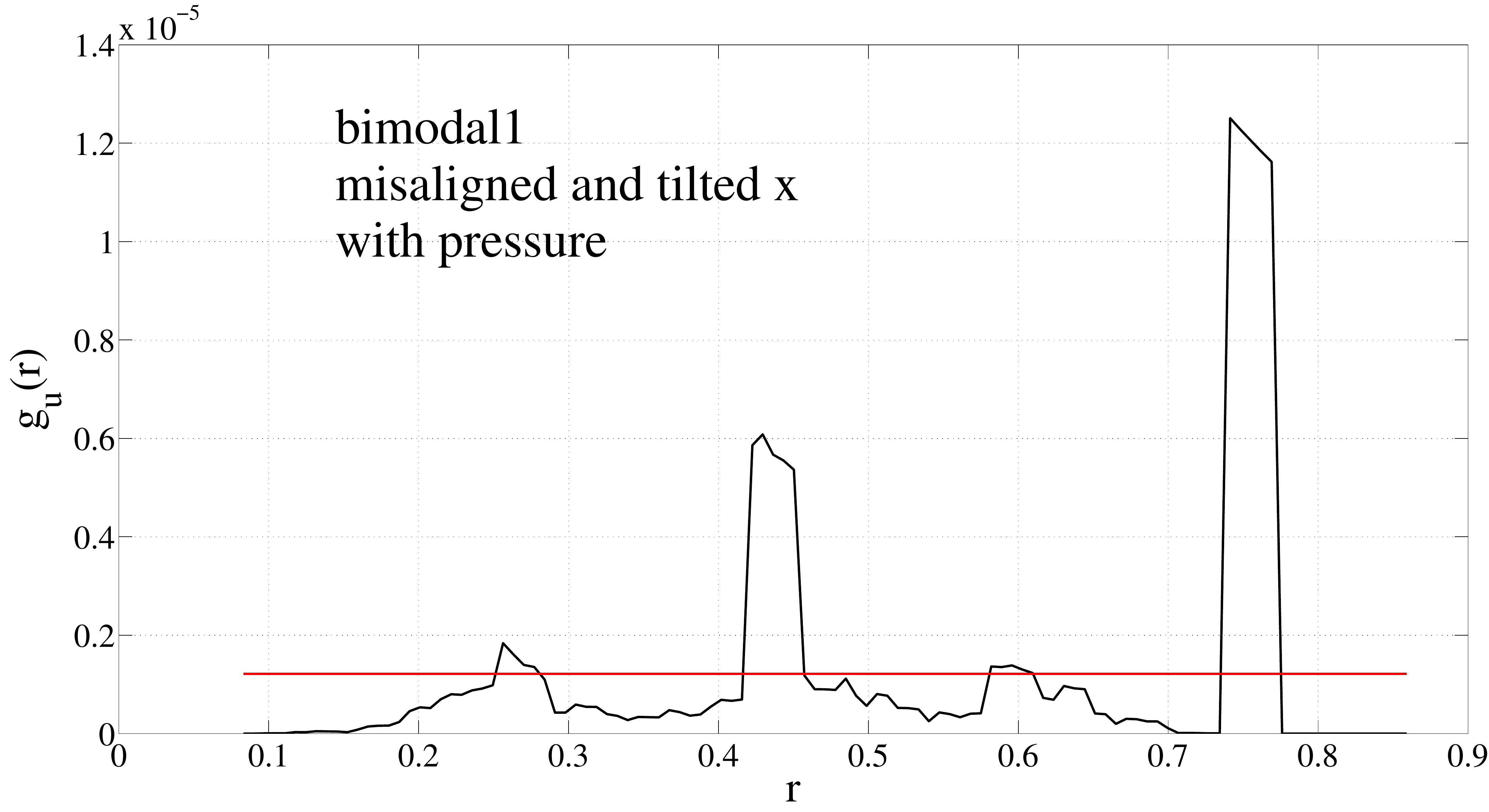}
\includegraphics[width=5.5cm]{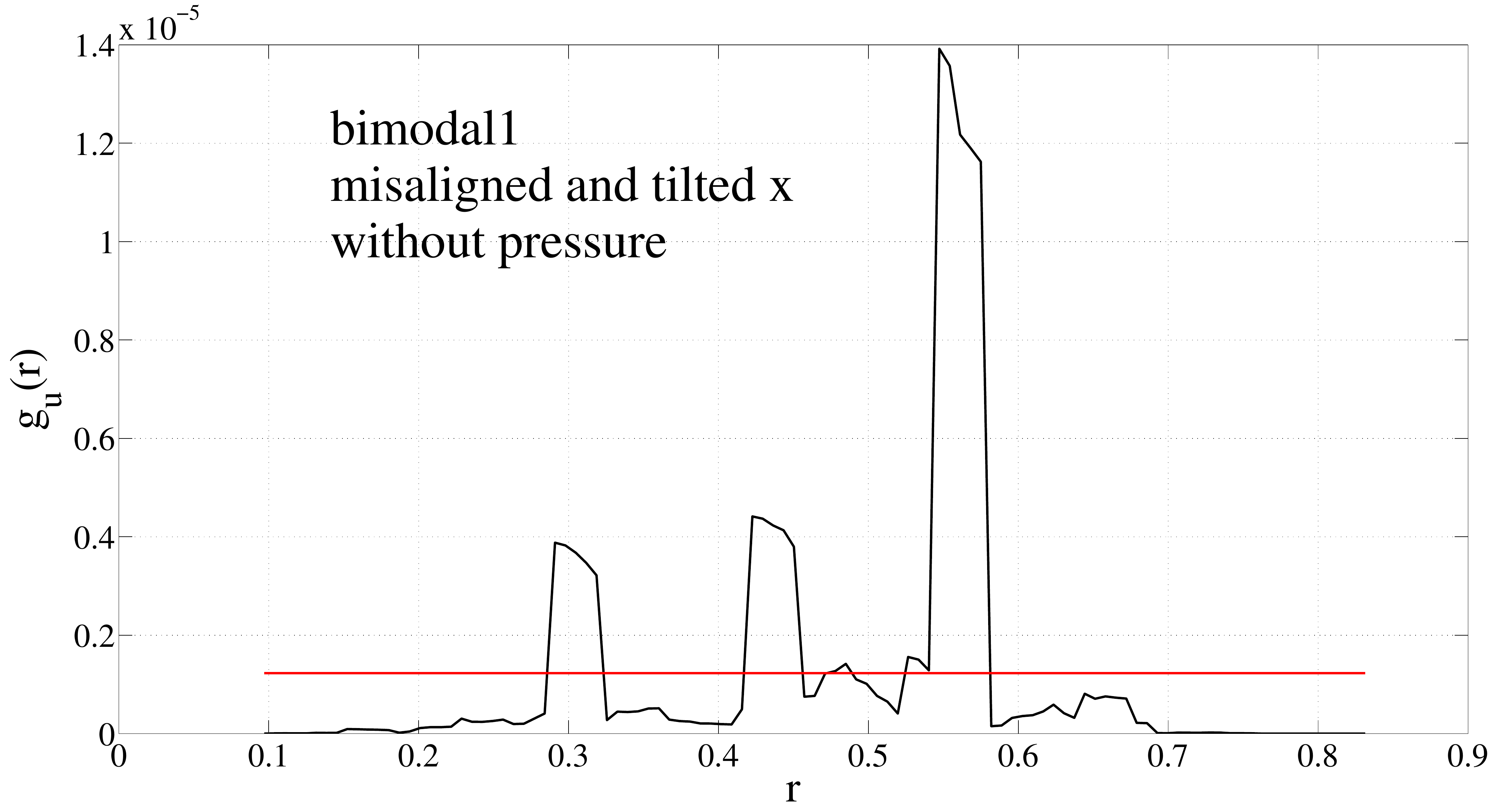}
\includegraphics[width=5.5cm]{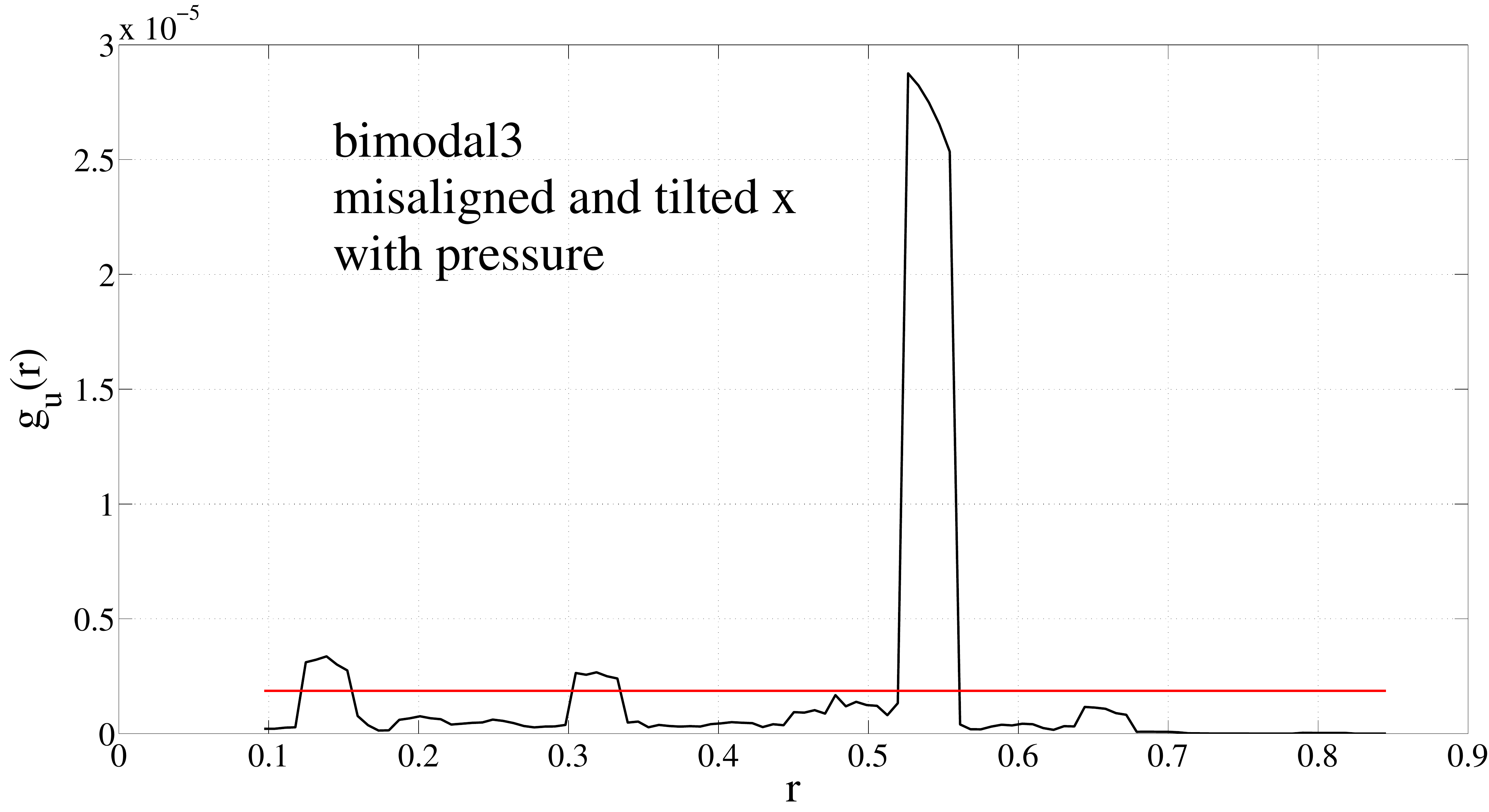}\\
\caption{Functions $g_{x}$, $g_{y}$, $g_{z}$ and $g_{u}$ for the mixtures bimodal1 and bimodal3 (cf.~Tab.~\ref{tab:powders}) 
for different orientational characteristics (see Tab.~\ref{tab:case}). $\Delta r$ is set to $5/256$ (cf.~Eq.~\ref{eq:rdf}).
For the powder bimodal1, small particles are omitted. The straight line in red shows the average value. It is calculated
by averaging the functions over the  bins.
\label{fig:corr}}
\end{figure*}

The emergence of orientational order can be addressed in a more specific way.
Let $B$ be the number of blocks. We define $B_{i}(r)$ as the ensemble of blocks inside 
a spherical shell of radius $r$ and width $\Delta r$ centered around  the $i$-th block. 
In order to take into account the local, relative orientation of pairs of blocks, we introduce the function
\begin{equation*}
S_{z}(r)=\sum_{i}^{B}\sum_{j\in B_{i}(r)}\frac{(\Gamma/2)^{2}}{[\theta_{z}(i)-\theta_{z}(j)]^{2}+(\Gamma/2)^{2}}\ .
\end{equation*}
$\theta_{z}(i)$ is the angle of rotation around the $z$ axis of the $i$-th block; the constant $\Gamma$
is assumed to be $10^{-4}$. In the above sum, the idea is that the contribution of every pair of 
block is "weighted'' by a Lorentz function. Summands close to $1$ indicate that the corresponding $\theta_{z}$
angles of the blocks differ slightly. Substitution into Eq.~\ref{eq:rdf} yields a kind of RDF
$g_{z}(r)$ measuring the average radial dependence of orientational correlations. Similar definitions
hold for the functions $g_{x}$ and $g_{y}$. For completeness, we also consider the function $g_{u}$ defined
by using the angle formed by the orientation vectors of pairs of blocks. For every block, the 
orientation vector is initially $\bm{u}=(1,0,0)$, to which three rotation transformations of angles
$\theta_{z}$, $\theta_{x}$ and $\theta_{y}$ are applied in this order. In Fig.~\ref{fig:corr}, we plot 
these functions for two powders in different cases of orientational randomness. For the functions 
$g_{z}$ of the bimodal1 powder (first row),
the peaks around $r=0.3$ indicate that the orientation of every block is to
a certain degree correlated with that of its adjacent blocks. From the further peaks
we can conclude that the orientation is also correlated between second-nearest 
neighbors (see Fig.~\ref{fig:layers}). 
For the functions $g_{x}$ and $g_{y}$ (second row), the average 
values are always at least a factor $20$ larger than for the previous $g_{z}$ functions.
The reason is that the tilt angles are restricted to values belonging to small intervals. 
The presence of more peaks is consistent with the elongated shape of particles.
For $g_{x}$ all peaks are comparable in height; instead, for $g_{y}$ there appears a marked peak around $r=0.3$.
Without pressure (third row), the peaks are reduced in height and that near $r=0.3$ is always the
more pronounced. This means that orientational correlations among second-nearest neighbors are
weaker because of looser packing. For the bimodal3 powder (fourth and fifth rows), from the peaks
in the range $r=0.1-0.2$ it turns out that out-of-plane blocks are also correlated. This peak
is of course higher for the functions $g_{x}$ and $g_{y}$. Furthermore, in this case the functions
$g_{x}$ and $g_{y}$ under the same conditions are similar (see second row). In the last row, the 
function $g_{u}$ is considered for three different powders. Direct comparison with the plots of the
functions $g_{x}$, $g_{y}$ and $g_{z}$ for the same type of powders indicate that the first peaks
flatten out. It follows that, for adjacent in-plane blocks, the angles $\theta_{x}$ are weakly
correlated; for out-of-plane blocks, it arises that the angles $\theta_{z}$ are instead weakly correlated.

We end this section by pointing out that every simulation used around $20$ MB of RAM memory and the
longest run took $74$h $43$min (bimodal3 powder aligned case with pressure) while the shortest one lasted for 
$3$h $51$min (fine powder tilted $x$ case without pressure). 
The program is written in Perl and it is executed on a desktop computer with an Intel processor of the family i7.
Finally, Tab.~\ref{tab:spheres} provides the packing fractions achieved with equivalent spherical fillers. These results are 
comparable with the predictions for random packing of spheres \cite{visscher,jodrey,chapin,spheres_princeton,williams4}, 
indicating that volume fractions can reach values in the interval $0.60$-$0.68$. It arises that the role of pressure is 
more effective with coarse blocks. The best volume fractions can not be reached within the present simulation settings because we 
can not tune arbitrarily parameters like the aspect ratios, the relative size between
small and large blocks and between the particles and the simulation domain, given the specific application under investigation. 
It clearly turns out that higher volume fractions can be realized only with more polydispersity.

\begin{table}[t]
\begin{center}
\begin{tabular}{l|ccccc}
\hline\hline
 & bimodal1 & bimodal2 & bimodal3 & granular & fine\\
\hline
no pressure   & $57.9\%$ & $53.2\%$ & $48.3\%$ & $38.5\%$ & $37.9\%$
\\
\hline
with pressure & $59.8\%$ & $54.8\%$ & $50.3\%$ & $41.6\%$ & $42.0\%$
\\
\hline\hline
\end{tabular}
\end{center}
\caption{\label{tab:spheres}
Packing fractions reached for powders composed of spherical particles. 
The average diameters of the two types of spheres are equal to the longest side of the blocks for
the corresponding mixture of Tab.~\ref{tab:powders}. Since simulations are much faster in this case,
smaller spheres are considered after $2\times 10^{6}$ unsuccessful attempts; 
all other simulation settings are maintained.}
\end{table}

\section{Conclusions}

It is well-known that the shape and size of the fillers affect the packing outcome and 
in turn, for example, the accurate prediction of the properties of materials by finite element methods \cite{garbo}. 
In this work we focused on small blocks because they better approximate the shape of graphitic 
clusters in some aspects (edges and corners). The algorithm is based on random sequential 
addition and can be extended easily to fillers with more irregular shapes, domains with 
solid walls or random addition mimicking deposition. In this approach, the interaction 
and evolution of the relevant constituents of the material are poorly accounted for. The 
main advantage of stochastic methods is the possibility to treat efficiently the complexity 
of the systems by concentrating on simple, effective rules \cite{kadanoff}. In the case of the present
algorithm, full control on interparticle distance seems to be the most useful property when the 
potential energy between hard particles is assumed to be identically zero, except for the case of contact. 
Then, coarser discretization clearly expedites the simulations.
Unavoidably, this leads to defects at the surfaces, but roughness is a ubiquitous 
element in materials science. In the case of bimodal packing our results show that more polydisperse
size distributions could lead to higher volume fractions, because particles are more suited to
fill voids comparable to their size. In any case, bimodal size distributions deserve attention
for the transport properties associated with their particular packing structures. Radial distribution
functions provides useful information on the size, amount and arrangement of the particles composing
the systems. Our analysis on correlations reveals that structural and orientational order do not
coincide. It turns out that a nonstructured medium (beyond the radial distance $r=0.3$ in our systems)
can still display forms of organization.

\section{Nomenclature}

\subsection{Symbols}

\begin{tabbing}
\hspace{1cm}=\hspace{0.5cm}\=\hspace{0.5cm}\=\kill
$B$\>-\>\begin{tabular}{p{6.4cm}}
number of block particles in the simulation domain
\end{tabular}\\
$B_{i}$\>-\>\begin{tabular}{p{6.4cm}}
number of blocks in a spherical shell of radius $r$ centered around the $i$-th block
\end{tabular}\\
$d_{\mathrm{min}}$\>-\>\begin{tabular}{p{6.4cm}}
minimal distance of approach between two block particles
\end{tabular}\\
$g_{\mathrm{ll}}$,$g_{\mathrm{ls}}$,$g_{\mathrm{ss}}$,$g$\>-\>\begin{tabular}{p{6.4cm}}
radial distribution functions
\end{tabular}\\
$g_{x}$,$g_{y}$,$g_{z}$,$g_{u}$\>-\>\begin{tabular}{p{6.4cm}}
functions defined for detecting orientational order
\end{tabular}\\
$l_{x}$,$l_{y}$,$l_{z}$\>-\>\begin{tabular}{p{6.4cm}}
side length of block particles
\end{tabular}\\
$N$\>-\>\begin{tabular}{p{6.4cm}}
number of voxels along each side of the simulation domain
\end{tabular}\\
$N_{1}$\>-\>\begin{tabular}{p{6.4cm}}
number of block particles of type $1$
\end{tabular}\\
$n_{x}$,$n_{y}$,$n_{z}$\>-\>\begin{tabular}{p{6.4cm}}
number of spheres composing each side of block particles
\end{tabular}\\
$P$\>-\>\begin{tabular}{p{6.4cm}}
number of pairs of blocks in the simulation domain
\end{tabular}\\
$p_{x}$,$p_{y}$,$p_{z}$\>-\>\begin{tabular}{p{6.4cm}}
probability for random moves along each axis
\end{tabular}\\
$r$\>-\>\begin{tabular}{p{6.4cm}}
radius of spheres composing block particles
\end{tabular}\\
$\Delta r$\>-\>\begin{tabular}{p{6.4cm}}
width of a spherical shell
\end{tabular}\\
$S$\>-\>\begin{tabular}{p{6.4cm}}
average number of block particles in a spherical shell of radius $r$
\end{tabular}\\
$S_{x}$,$S_{y}$,$S_{z}$,$S_{u}$\>-\>\begin{tabular}{p{6.4cm}}
weighted number of block particles in a spherical shell of radius $r$
\end{tabular}\\
$t$\>s\>\begin{tabular}{p{6.4cm}}
time
\end{tabular}\\
$\bm{u}$\>-\>\begin{tabular}{p{6.4cm}}
unit vector
\end{tabular}\\
$x$,$y$,$z$\>-\>\begin{tabular}{p{6.4cm}}
cartesian coordinates
\end{tabular}\\
$V$\>-\>\begin{tabular}{p{6.4cm}}
volume of simulation domain
\end{tabular}\\
$\Gamma$\>-\>\begin{tabular}{p{6.4cm}}
model parameter
\end{tabular}\\
$\gamma$\>-\>\begin{tabular}{p{6.4cm}}
power-law exponent
\end{tabular}\\
$\phi$\>-\>\begin{tabular}{p{6.4cm}}
volume fraction
\end{tabular}\\
$\sigma$\>-\>\begin{tabular}{p{6.4cm}}
standard deviation of a Gauss distribution
\end{tabular}\\
$\theta_{x}$,$\theta_{y}$,$\theta_{z}$\>-\>\begin{tabular}{p{6.4cm}}
rotation angles around each axis
\end{tabular}\\
\end{tabbing}

\subsection{Abbreviations}

\begin{tabbing}
\hspace{1cm}=\hspace{0.5cm}\=\hspace{0.5cm}\=\kill
min\>\>minimum\\
part.\>\>particles\\
press\>\>with pressure\\
RDF\>\>radial distribution function\\
RSA\>\>random sequential addition\\
std\>\>standard\\
\end{tabbing}

\begin{acknowledgments}
This work was supported through project BiPCaNP (P.~No.~10055.1)
by the Swiss Commission for Technological Innovation (KTI/CTI).
\end{acknowledgments}



\begin{thebibliography}{1}
\bibitem{note}This choice is tightly related to the way the blocks are obtained from
the spheres (cf.~next paragraph). The distance $d_{\mathrm{min}}$ is introduced
so that the blocks remain completely impenetrable, i.e., with no contact
between their surfaces.
\end{thebibliography}
\end{document}